\documentclass{aa}
\usepackage{natbib}
\bibpunct{(}{)}{;}{a}{}{,} 
\usepackage{graphicx}
\usepackage[colorlinks=true,linkcolor=blue,urlcolor=blue,citecolor=blue]{hyperref}
\usepackage{txfonts}
\usepackage{mathtools}
\usepackage{amsmath}
\usepackage{multirow}
\usepackage{placeins}
\makeatletter
\renewcommand*\aa@pageof{, page \thepage{} of \pageref*{LastPage}}
\makeatother
\usepackage{xcolor}

\begin{document}

   \title{Characterisation of local halo building blocks: Thamnos and Sequoia}

   \author{Emma Dodd\inst{1}
   \and Tom\'{a}s Ruiz-Lara\inst{2,3}
   \and Amina Helmi\inst{1}
   \and Carme Gallart \inst{4,5} 
   \and Thomas M. Callingham\inst{1}
   \and Santi Cassisi \inst{6,7}
   \and Emma Fern\'{a}ndez-Alvar \inst{4,5}
   \and Francisco Surot \inst{4} 
}

   \institute{
      Kapteyn Astronomical Institute, University of Groningen, Landleven 12, 9747 AD Groningen, The Netherlands\relax
       \and Universidad de Granada, Departamento de Física Teórica y del Cosmos, Campus Fuente Nueva, Edificio Mecenas, E-18071, Granada, Spain\relax
       \and Instituto Carlos I de Física Te\'orica y computacional, Universidad de Granada, E-18071 Granada, Spain\relax
       \and Instituto de Astrofísica de Canarias, E-38200 La Laguna, Tenerife, Spain \relax
       \and Departamento de Astrofísica, Universidad de La Laguna, E-38205 La Laguna, Tenerife, Spain \relax
        \and INAF – Astronomical Observatory of Abruzzo, via M. Maggini, sn, 64100 Teramo, Italy \relax
        \and INFN, Sezione di Pisa, Largo Pontecorvo 3, 56127 Pisa, Italy
             }

   \date{}

 \abstract
   {A crucial aspect of galaxy evolution is the pace at which galaxies build up their mass. We can investigate this hierarchical assembly by uncovering and timing accretion events experienced by our Galaxy.}
   {In the Milky Way, accreted debris has been previously identified in the local halo, thanks to the advent of \textit{Gaia} data. We aim to couple this dataset with advancements in colour-magnitude diagram (CMD) fitting techniques to characterise the building blocks of the Galaxy, based on their age and metallicity distributions. Here, we focus on the retrograde halo, specifically those of Thamnos and Sequoia. }
   {We conducted this study as part of the ChronoGal project by fitting the absolute CMDs (using CMDft.\textit{Gaia}) of samples of stars associated with these sub-structures, extracted from a local 5D \textit{Gaia} DR3 dataset. 
   Comparing their derived age and metallicity distributions with those of the expected contamination, from the dominant \textit{Gaia} Enceladus (GE) and low-energy (LE) in situ populations, we can unveil the stellar population signatures of the progenitors of Sequoia and Thamnos. }
   {We show that both Thamnos and Sequoia have a metal-poor population ([Fe/H] $\sim -2.5$ to $-1.5$ dex) that is distinct from the expected contamination. Their age distributions offer us the ability to see the pace of the build-up of their progenitors. Half of the stars in Sequoia were formed by the look-back time of $ \sim 12 ^{+0.3}_{-0.3}$ Gyr. 
   Thamnos appears slightly older, on average, and declines quickly, having formed half its stars at $\sim 12.3 ^{+0.3}_{-0.3}$ Gyr. Compared to GE and the LE in situ populations, they formed half of their stars by $12.1 ^{+0.1}_{-0.1}$ Gyr and $12.9 ^{+0.1}_{-0.1}$ Gyr, respectively. Caution should be taken when interpreting the age distributions, especially that of Sequoia, due to the low number of stars, which can cause shifts to younger ages of up to $\sim 1$ Gyr. However, considering these potential shifts and the underlying contamination that is inherently difficult to remove completely, our results allow us to safely conclude that Thamnos, \textit{Gaia} Enceladus, and Sequoia are all predominantly old and accreted at similar epochs, within $\sim$ 1-2 Gyr of each other. }
   {We  present, for the first time, the age distributions of the retrograde halo sub-structures: Sequoia and Thamnos. These have been derived from purely photometric data using CMD fitting techniques, which also provide metallicity distributions that successfully reproduce the results from spectroscopy, highlighting the applicability of CMDft.\textit{Gaia}.}


   \maketitle
%

\section{Introduction}

According to $\Lambda$CDM cosmology, galaxies assemble hierarchically over time, growing their halos through mergers and the accretion of smaller galaxies \citep{springel_simulations_2005}. Identifying and characterising the debris in our own Galaxy's halo offers insights into the significance of accretion in this process, along with  overall processes of galaxy formation and evolution.

Thanks to the advent of \textit{Gaia} data \citep{prusti2016gaia,GaiaDR3_Summary}, we are now able to identify this type of merger debris in our local stellar halo as an over-density in the integrals of motion (IoM) space \citep{helmi2000mapping}. This has provided insights into the assembly history, as well as on how mergers have played a role in shaping the Galaxy 
\citep[for recent reviews see][]{helmi_streams_20,deason2024galactic}. 
\textit{Gaia} data have revealed that the local stellar halo is dominated by debris from a single merger, \textit{Gaia} Enceladus/Sausage \citep{helmi_merger_2018,belokurov2018}. Since its discovery, extensive work has been done to characterise the debris of this merger, 
providing stellar mass estimates in the range of $\sim$ 10$^8$-10$^{10}$ M$_{\odot}$ \citep[e.g.][ with more recent studies predicting toward the lower mass end than initial estimates]{helmi_merger_2018,vincenzo2019,feuillet2020,mackereth2020weighing,lane2023stellar} and an estimated accretion time of $\sim$ 10 Gyr ago, equivalently $z\sim 2$ \citep[e.g.][]{helmi_merger_2018,gallart2019uncovering,bonaca2020timing,montalban2021chronologically}. 
Additionally, several other dynamical sub-structures have been uncovered \citep[e.g.][]{koppelman2019multiple,naidu_evidence_2020, yuan2020, Horta2021,lovdal2022,ruizlara2022, Dodd2023}, although their nature (accreted or in situ) and  characterisation are rather limited or even non-existent. 

The retrograde part of the halo has been revealed to be especially rich in small sub-structures that are either chemically or kinematically distinct from one another \citep[e.g.][]{Dodd2023,Oria_Anaeus_22_2022arXiv220610404O,ceccarelli2024walk}. The dominant debris appears to come from three main accretion events; \textit{Gaia} Enceladus, Sequoia, and Thamnos, the latter of which have not yet been characterised (in detail).

Sequoia is a loosely bound retrograde sub-structure identified first by \citet{myeong2019} as a chemically distinct accretion event in the halo \citep[see also][]{matsuno2019origin,matsuno2022}. An estimate of the stellar mass of $\sim5\times10^7$ M$_\odot$ has been proposed,  along with an accretion time of around 9 Gyr ago \citep{myeong2019}. However, it is becoming more clear that the retrograde halo is more complex than we first thought. 
Simulations have shown that debris from a massive merger such as \textit{Gaia} Enceladus (GE, hereafter) can populate this region, depending on its initial orbital configuration and morphology \citep[e.g.][]{Koppelman2020,naidu2021reconstructing,amarante2022}. Observations have revealed that even though Sequoia stars at low metallicity  are chemically distinct from GE \citep{myeong2019,matsuno2022}, above [Fe/H] $\sim$ $-1.5,$ the chemical abundances are indistinguishable from those of GE debris \citep{ceccarelli2024walk}. The level of contamination by GE in that region of IoM is suggested to be $\sim$10-20\% \citep{matsuno2022}.

Initially, Sequoia was defined in \citet{myeong2019} to extend in energy down to low orbital energies (high binding energy). 
Since then, \citet{koppelman2019multiple} discovered the Thamnos sub-structure, separating the retrograde halo into high $E_n$ (Sequoia) and low $E_n$ (Thamnos) for the first time, owing to differences in chemistry between the two populations. Given the mean metallicities of the Sequoia stars, it seems unlikely that its progenitor could have been massive enough \citep[because of the mass-metallicity relationship; e.g.][]{kirby2013universal,ma2016origin} to create such a large spread in IoM, as that first proposed by \citet{myeong2019}. 

Thamnos itself was first defined by  \citet{koppelman2019multiple} as two independent sub-structures: Thamnos 1 and 2. Given their similar abundances and stellar populations, it is possible that the two sub-structures share the same progenitor \citep[see also][]{ruizlara2022} and from the extent in IoM space, \citet{koppelman2019multiple} placed an upper limit on the stellar mass of $5\times10^6$ M$_{\odot}$. 
It is worth noting that Thamnos 2, which is the less retrograde and more bound sub-structure, 
contains stars that have [Fe/H] of $\sim -1.3$ dex that are Al-enhanced; a population that is likely to be contamination of in situ origin (see Fig. 4 in \citealt{koppelman2019multiple}, and also \citealt{Horta2021}). Subsequently, \citet{ruizlara2022} using {\it Gaia} EDR3 supplemented by LAMOST \citep{cui2012}, also confirmed that Thamnos 2 is more metal-rich than Thamnos 1, again possibly due to increasing contamination as $L_z$ becomes less negative. We note, however, that this is based on a limited number of stars with chemical information. 
More recently, thanks to \textit{\textit{Gaia}} DR3 data, Thamnos has been identified with clustering in IoM space as one single sub-structure \citep{Dodd2023} and its chemistry still displays a mix of stellar populations, including contamination from in situ components and probably from \textit{Gaia} Enceladus as well. 

In addition to accreted sub-structures, the local halo is known to host structures of an in situ origin. The colour-magnitude diagram (CMD) of local halo stars was shown with \textit{Gaia} DR2 to consist of two sequences \citep{babusiaux2018gaia}. The blue sequence has been attributed to the accreted halo and dominated by \textit{Gaia} Enceladus debris and the red sequence is said to be heated thick-disc stars \citep[e.g.][]{haywood2018disguise}. It is now believed that thick-disc stars already present during the merger with \textit{Gaia} Enceladus were heated onto halo-like orbits, 
producing what we now call the hot thick disc or `the Splash' \citep[e.g. see][]{helmi_merger_2018,gallart2019uncovering,belokurov2020splash}.

A more ancient in situ halo component has been uncovered using [Al/Fe] abundances, referred to as the Aurora population \citep{belokurov2022dawn}, and is hypothesised to have formed during a period of chaotic (pre-disc) evolution. 
This population exhibits a rapid evolution of [Al/Fe] to anomalously high [Al/Fe], which can be explained by a rapidly star-forming and self-enriching galaxy before supernovae of type Ia (SN Ia) start to contribute significantly and dilute the [Al/Fe]. As star formation happens on a longer timescale in dwarf galaxies, they do not reach as high [Al/Fe] before the SN Ia kick in \citep{hawkins2015using,belokurov2022dawn}. The chemical abundances of the Aurora population also suggest that the proto-Milky Way formed with a large contribution from massive star clusters \citep{belokurov2022dawn, myeong2022milky}. However, \citet{Horta2021,horta2024proto} show that this proto-Milky Way population can be chemically indistinguishable from an early (massive) merger~event.

Our understanding of the accretion events that contribute to the local Milky Way halo has, to date, mostly focussed on a dynamical perspective, made possible thanks to the \textit{Gaia} mission. For sub-sets of the stars associated to sub-structures, this has been coupled with chemical information from ground-based spectroscopic surveys (e.g. LAMOST, \citealt{zhao2012lamost}; GALAH, \citealt{de2015galah}; APOGEE, \citealt{majewski2017apache}) or targeted follow-up \citep[e.g.][]{aguado2021,matsuno2022,ceccarelli2024walk} to obtain a chemo-dynamical view.   
The future of the characterisation of halo sub-structures relies upon our access to the chemistry and age information    for a larger sample of stars, for which we already have the dynamical information from \textit{Gaia}. The chemical characterisation will improve greatly with the upcoming wide-area spectroscopic surveys, such as WEAVE \citep{jin2023wide} and 4MOST \citep{de20194most}. 

An important dimension of the assembly history that is still (for the most part) missing is the chronology of the individual accretion events. Specifically, to be able to assess how quickly galaxies build up their stellar populations, as well as to determine what the timelines for galaxy assembly and the duration of merger events are.
Unfortunately, age estimates for stars are notoriously difficult to obtain for large samples, since the ages of stars cannot be directly measured and have to be inferred by comparing observed properties with stellar evolution models.  
Currently, there are two main methods for obtaining age estimates of individual stars: isochrone fitting and asteroseismology \citep[e.g. for a review see][]{soderblom2010ages}; both of which have their own limitations in terms of precision and sample sizes.

Alternatively, stellar age distributions can be obtained from a CMD fitting of composite stellar populations \citep{dolphin2002numerical,cignoni2010star}.
This technique has already been well established for deriving the star formation histories of dwarf satellites in the Local Group \citep[e.g.][]{gallart1999star, noel2009old,tolstoy2009, bernard2012star,weisz2012star, cignoni2013,cole2014,skillman2017islands,mcquinn2024jwst}. 
Thanks to the combination of \textit{Gaia} data and developments in the CMD fitting technique, it is now feasible to do this for homogeneous samples of stars in the Milky Way \citep[e.g.][]{ruiz2020recurrent,dal2021dissecting,mazzi2024dissecting,gallart2024}, including the building blocks identified as accreted sub-structures (e.g. see \citealt{gallart2019uncovering} and \citealt{ruiz2022_HS_SFH} for the characterisation of the stellar content of \textit{Gaia} Enceladus and the Helmi streams, respectively). 
We note that this method does not require any assumptions on the age-metallicity relation, the metallicity distributions, or the functional form of the star formation rate over time (for more details, see \citealt{gallart2024}). It provides us with a detailed age distribution for a given sample of stars. Applying this to the building blocks of the Milky Way allows us to construct a timeline of the mergers our Galaxy has experienced and examining when a system had been quenched by accretion.

In this paper, we focus on Thamnos and Sequoia. By fitting their absolute CMDs, we have been able to obtain age and metallicity distributions for these sub-structures. Section \ref{sec:data} presents the datasets used in this study and the selections of our samples. The methodology is presented in Sect. \ref{sec:method}, including the CMD fitting procedure (Sect. \ref{sec:fitting_CMD}), the treatment of contamination (Sect. \ref{sec:contaminants}), and the performance of mock tests to validate the results (Sect. \ref{sec:mocks}). The results are presented in Sect. \ref{sec:results} were derived using only photometry, demonstrating that the metallicity
distributions of the sub-structures resemble the spectroscopic distributions,  while also providing  the added age dimension that enables us to see more clearly differences in their formation histories. A discussion of these results is presented in Sect. \ref{sec:discussion}, highlighting the importance of taking into account contamination from the dominant \textit{Gaia} Enceladus and the in situ halo populations (Sect. \ref{sec:discussion_contamination}) and the complexity of the Thamnos region of IoM space (Sect. \ref{sec:Th_selection}). 
Section \ref{sec:conclusion} summarises our results and conclusions.

\begin{figure*}[h]
\centering
\includegraphics[width=\textwidth]{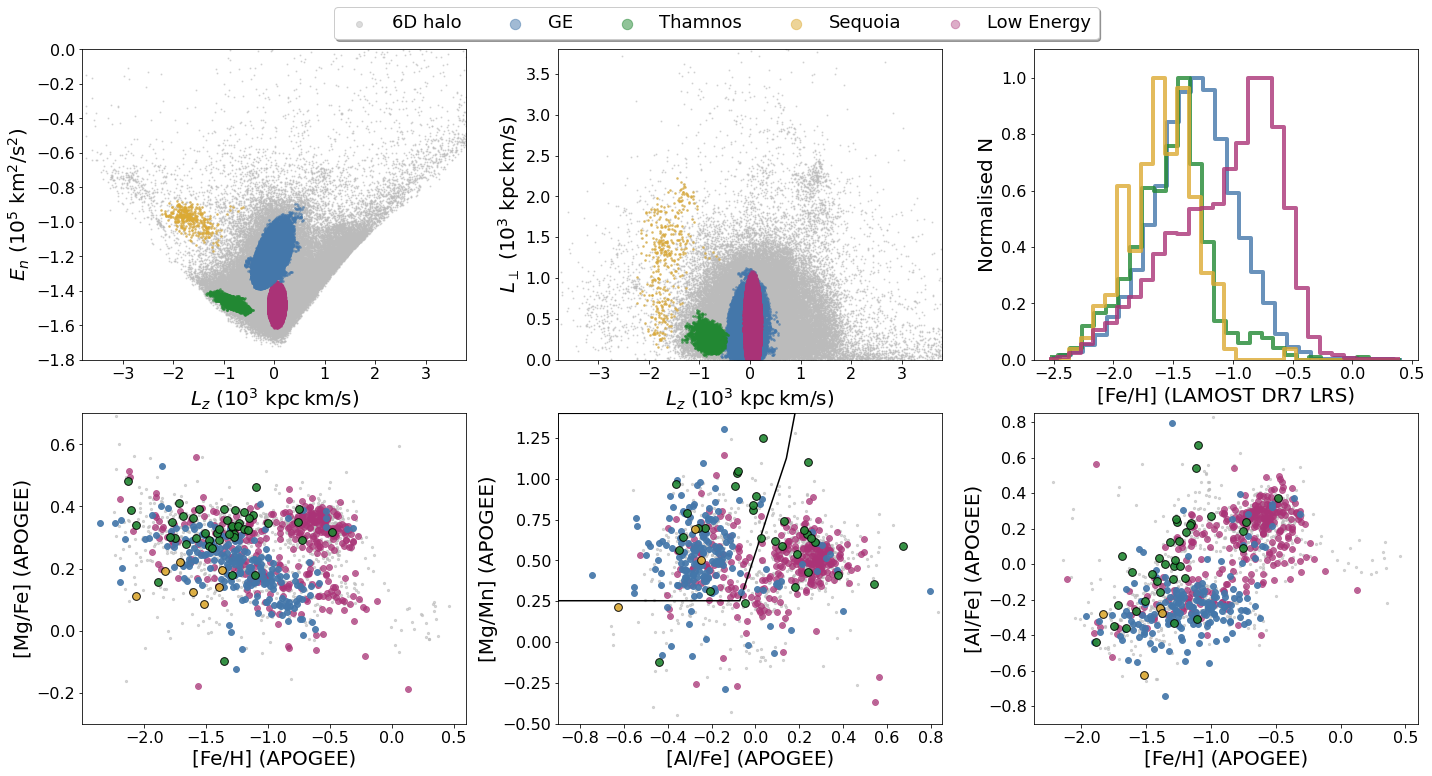}
\caption{6D samples relevant to this work. We selected \textit{Gaia} Enceladus, Thamnos, and Sequoia, according to \citet{Dodd2023} and defined another sample referred to as the low-energy (LE) sample, shown in magenta. This sample mimics the contamination we expect at low energy for Thamnos, as we can see more clearly in the APOGEE DR17 abundances (bottom row). The top row shows integrals of motion (IoM) space and LAMOST DR7 LRS metallicity distributions normalised and binned each with a relative shift of 0.01 dex to aid in the comparison. The bottom row shows APOGEE DR17 abundances for the different sub-structures, where the Thamnos and Sequoia stars are shown with a black outline. }
\label{6D_characterisation}
\end{figure*}

\section{Data} \label{sec:data}
In a previous work, we developed a clustering algorithm that is able to identify several overdensitites, including Thamnos and Sequoia, when applied
to the 6D local halo sample in a statistically
robust manner  \citep{lovdal2022,ruizlara2022,Dodd2023}. 
To perform the CMD fitting, we must be able to model the selection functions of our samples; so, for our 6D samples this means the \textit{Gaia} RVS selection function 
(see Sect.~\ref{sec:dispar}). For smaller sub-structures (e.g. Sequoia), the number of member stars with radial velocities remains limited, making CMD fitting of these 6D samples difficult.
For these reasons, we turned to 5D datasets. We first used the 6D samples to select the sub-structures (Sec~\ref{sec:6D_data} \& \ref{sec:6D_subs}) and to then motivate our selections from a local 5D sample (Sec~\ref{sec:5D}; similarly to what was done in \citealt{ruiz2022_HS_SFH}).
Selecting our samples in 5D allows us to avoid having to model complicated selection functions. Even more importantly, it increases the number of bright stars, allowing us to obtain a complete sample down to $M_G < 5$ within 2.5~kpc.

\subsection{6D Halo Sample}\label{sec:6D_data}
Our 6D halo sample was defined from \textit{\textit{Gaia}} DR3 data according to \citet{Dodd2023}. We selected stars with (total) relative parallax uncertainty less than 20\% as: $(\varpi -\Delta_\varpi) /\sqrt{\sigma_{\textrm{parallax}}^2 + \sigma_{\textrm{sys}}^2} \ge 5 $,
where $\Delta_\varpi$ represents the individual zero-point offsets determined following \citet{lindegren2021}, $\sigma_{\textrm{parallax}}$ is the \texttt{parallax\_error}, 
and $\sigma_{\textrm{sys}}$ is the systematic uncertainty on the zero-point, which we take to be 0.015 mas  \citep{lindegren2021}. To obtain a distance, we inverted the corrected parallax and we selected stars within 2.5~kpc of the Sun. One difference  with respect to the study of \citet{Dodd2023} is that we here we have extended our sample from stars with a radial velocity in \textit{\textit{Gaia}} DR3 to also include  stars with a radial velocity from ground-based spectroscopic surveys (as done also for EDR3 in \citealt{lovdal2022}) such as LAMOST DR7 \citep{cui2012}, APOGEE DR17 \citep{accetta2022}, GALAH DR3
\citep{Buder2020}, and RAVE DR6 \citep{steinmetz2020}. The substructures are defined still as those in \citet{Dodd2023} from clustering on the RVS sample,  which is more robust.  Then we added extra stars with radial velocities from ground-based spectroscopic surveys to our predefined clusters (see Sect.  \ref{sec:6D_subs} for more details).

We cleaned the sample with \texttt{RUWE} $< 1.4$, line of sight velocity uncertainty, $\epsilon\,(V_{\rm los}\footnote{after applying the correction to \texttt{radial\_velocity\_error} recommended by  \citet{Babusiaux2022}}) < 20$~km/s, \texttt{(G$_{RVS}$ - G)}$> -3$\footnote{to remove problematic radial velocities as recommended by  \citet{Babusiaux2022}}, and for stars with $|b| < 7.5^{\rm o}$ we require\footnote{to avoid highly contaminated spectra and spurious velocities, following
\citet[][Sect.~9]{Katz2022}} \verb|rv_expected_sig_to_noise|~$>5$.
After correcting for the solar motion using 
$ (U, V, W )_\odot$ = (11.1, 12.24, 7.25) km/s \citep{schonrich2010}
and for the motion of the local standard of rest (LSR) using a $|\textbf{V}_{\textrm{LSR}}|$ of 232.8\,km/s \citep{mcmillan2017}, we selected halo stars as |$\textbf{V}-\textbf{V}_\textrm{LSR}$| $>$ 210 km/s. 
This resulted in a 6D halo sample of 101,419 stars, 71,849 of which have a \textit{Gaia} radial velocity.
We calculate the integrals of motion (IoM) for stars (energy; $E$, and angular momenta; $L_z$, $L_{\perp}$), computing $E$ using the same potential as in  \citet{Dodd2023}. This potential consists of a Miyamoto-Nagai disc with parameters $(a_d, b_d) = (6.5, 0.26)$ kpc, $M_{d}=9.3\times 10^{10} M_\odot$,
a Hernquist bulge with $c_b = 0.7$ kpc, $M_{b}=3.0 \times 10^{10} M_\odot$, 
and an NFW halo with  $r_s=21.5$ kpc, $c_h$=12, and $M_{\rm halo}=10^{12} M_\odot$.
We defined 
$L_z$ to be positive for prograde stars. While we limited our sample to those with $\epsilon\,(V_{\rm los}) < 20$~km/s, the majority of the stars have a significantly smaller $V_{\rm los}$ uncertainty. The mean $V_{\rm los}$ uncertainty for the full sample is 5.2~km/s and for the \textit{Gaia} RVS sub-sample is 4.5~km/s.
By propagating the errors in $V_{\textrm{los}}$ into errors in IoM space ($E$, $L_z$, $L_{\perp}$), we  checked that these uncertainties had limited effect on the assigning of stars to sub-structures.

\subsection{Sub-structures in 6D}\label{sec:6D_subs}
 \begin{figure}[h]
\centering
\includegraphics[width=0.47\textwidth]{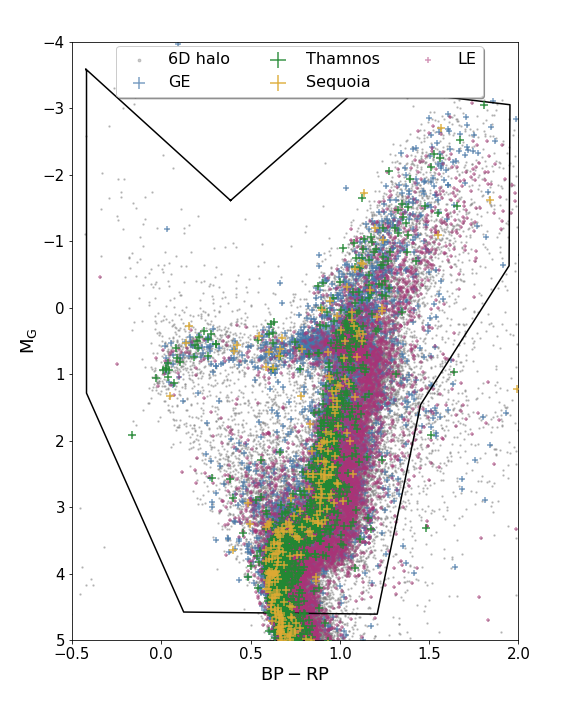}
\caption{CMD of bright ($M_G <5$) 6D selected Thamnos, Sequoia, \textit{Gaia} Enceladus (GE), and low energy (LE) samples. The region outlined in black is the bundle used for fitting the CMDs, as described in Sect.~\ref{sec:fitting_CMD}, and is chosen to extend $\approx$1 magnitude below the main sequence turn-off. The background sample in grey is our full 6D halo sample. }
\label{fig:CMD_6D}
\end{figure}

We followed the definitions for the local halo sub-structures identified in \citet{Dodd2023}, obtained by clustering on the \textit{Gaia} DR3 RVS sample that has lower uncertainties. Then we added extra stars with radial velocities from ground-based spectroscopic surveys to our predefined sub-structures, by selecting those that fall within an ellipsoid which contains 80\% of the original sub-structure members. This is equivalent to cutting in Mahalanobis distance\footnote{The Mahalanobis distance between a star from a ground-based spectroscopic survey and a sub-structure in IoM space is given as: 
$D^{\prime} = \sqrt{({\boldsymbol \mu}_{\textrm{sub}}-{\boldsymbol \mu}_{\textrm{*}})^T \Sigma_{\textrm{sub}}^{-1} ({\boldsymbol \mu}_{\textrm{sub}}-{\boldsymbol \mu}_{\textrm{*}})}$
where $\boldsymbol{\mu}_{\textrm{sub}}$ and $\Sigma_{\textrm{sub}}$  correspond to the mean and covariance matrix of the sub-structure stars, and $\boldsymbol{\mu}_{\textrm{*}}$ is the position of the star in IoM space.} at 2.13, which was found previously to minimise adding noise to sub-structures \citep[see][for more details]{ruizlara2022}.

Figure~\ref{6D_characterisation} shows this 6D halo sample within 2.5 kpc, with \textit{Gaia} Enceladus (GE), Thamnos, and Sequoia selections shown in coloured markers. We also included another sub-set that is seen to be below GE in terms of energy,  which we refer to as the `low energy' (LE) sample. This samples was selected to mimic the contamination we expect in the Thamnos region of IoM space.  
This selection is defined using stars in a narrow range of $v_y$ between $-230$ and $-250$ km/s, and within $|v_x| < 180$ km/s, combined with a cut of $E_n < -1.35 \times 10^5$ km$^2$/s$^2$ so that we are not overlapping with our pure GE selection in energy. This selects a general halo sample on highly bound orbits and appears to be a mix of populations, including \textit{Gaia} Enceladus debris at low energy and with an in situ component.  

Figure~\ref{fig:CMD_6D} shows the CMDs of these populations, 
demonstrating that all of these sub-structures are composed mainly of old stars with metallicities that are different (as shown in Fig.~\ref{6D_characterisation}) among the structures (as seen by the different colours of the RGB and position of the old main sequence turn-off; oMSTO). The CMD fitting will help us further unveil these differences in age and metallicity. We note that both Sequoia and Thamnos exhibit an extended horizontal branch with higher relative contributions of horizontal branch (HB) stars towards the blue end (especially in Thamnos; see Fig.~\ref{fig:CMD_6D}).

Figure~\ref{6D_characterisation} shows that Sequoia is a sub-structure lying in a region of  IoM space that is relatively well separated from other sub-structures. We define Sequoia as a smaller region in IoM space than the typical size assumed  in past studies \citep[e.g.][for comparison see Figure 13 in \citealt{lovdal2022}]{myeong2019, koppelman2019multiple, yuan2020, naidu_evidence_2020}. This is because \citet{ruizlara2022} showed that the larger retrograde selection includes sub-structures that have distinct kinematics and that given Sequoia's estimated mass (from its mean metallicity), it would be inconsistent with a very broad extent in the IoM space. 
As previously outlined, it is evident from both simulations \citep[e.g.][]{Koppelman2020,naidu2021reconstructing,amarante2022} and observations using chemical abundances \citep[e.g.][]{matsuno2022,ceccarelli2024walk} that this region of IoM space is contaminated by debris from the \textit{Gaia} Enceladus merger. This is something that we consider in this work. 

Thamnos is a low (orbital) energy retrograde structure (see Fig.~\ref{6D_characterisation}) and its location at low energy means that any selection of Thamnos stars will be contaminated from \textit{Gaia} Enceladus \citep{helmi_merger_2018} and other populations, such as possibly an in situ halo or Aurora (i.e. the proto-Galaxy), as both dominate at low energy \citep{belokurov2022dawn}. These structures are much larger than what we believe the true Thamnos population to be and, hence, on average more metal-rich. This is why we introduced the LE sample, which is a combination of the in situ halo and likely some \textit{Gaia} Enceladus debris,  used to decontaminate Thamnos and characterise the true signal in our analyses.

\subsubsection{Metallicity distributions}
To  quantitatively establish the differences in the LAMOST DR7 LRS metallicity distributions of Sequoia with GE and Thamnos with the LE sample, we compared the MDFs using the two sample Kolmogorov–Smirnov (KS) test \citep{kolmogrov1933,smirnov1939estimate}, where the KS statistic gives the maximum difference of the cumulative metallicity distributions. 

For Sequoia and \textit{Gaia} Enceladus, we found a KS test statistic of 0.38 (with a significance of $\sim 9 \sigma$) and we can confidently say that the two samples are not drawn from the same distribution. This maximum difference in the cumulative metallicity distributions occurs at [Fe/H] = $-1.31$ dex, however, the distributions are already statistically different at the 95\% level\footnote{using that the critical value for the two sample KS test of sizes $n_1, n_2$ can be approximated as $c(\alpha) \sqrt{(n_1 + n_2)/n_1\,n_2}$ with $c(0.05)$ = 1.36} from [Fe/H]~$\sim$~$-$1.8.

For Thamnos (looking at Fig.~\ref{6D_characterisation}), we can see from the metallicity distributions that Thamnos (green) has a more pronounced metal-poor tail around [Fe/H] $\lesssim -2$ than GE and the LE sample, which we believe is the signature of the small dwarf galaxy. On the other hand, the more metal-rich stars in the Thamnos region are a signature of this contamination. In fact, Fig.~\ref{6D_characterisation} shows that there is a large peak in the metallicity distribution around [Fe/H] $\sim -1.4$ dex, which is a combination of GE contamination and an Al-enhanced population of similar metallicity that can be seen in the bottom-left and right-most panels. This is the signature of the ancient population of Aurora \citep{belokurov2022dawn} and likely the result of intense rapid star formation in the early Milky Way. 

By statistically comparing the metallicity distributions for Thamnos and our LE  sample, we can confidently say that the samples are drawn from different distributions, with a KS test statistic of 0.54 (at a significance of $> 10 \sigma$). 
This maximum difference of the cumulative metallicity distributions occurs at [Fe/H] $= -1.13$ dex; however, the distributions are already statistically different at the 95\% level from [Fe/H] $\sim$ $-1.9$ dex. This supports that there is a statistically significant metal-poor tail in our Thamnos selection corresponding to something other than the overall low-energy halo.

\subsubsection{Chemical abundances}
In the bottom row of Fig.~\ref{6D_characterisation} we show APOGEE DR17 abundances of the 6D sub-structures relevant to this work, which supports our definitions of contamination. We show in the first panel the [$\alpha$/Fe]-[Fe/H] plane with Mg as our $\alpha$. Accreted galaxies are expected to have lower [$\alpha$/Fe] at fixed [Fe/H] than in situ stars since this is indicative of a slower enrichment \citep[e.g.][]{horta2023chemical}. Individual progenitors are expected to follow distinct tracks in this space; for example, Sequoia which has been shown to present lower [$\alpha$/Fe] than GE at fixed metallicity \citep{matsuno2022}.
The middle panel shows the [Al/Fe]--[Mg/Mn] plane, which is commonly used to distinguish accreted (chemically unevolved) and in-situ (chemically evolved) populations \citep[as proposed by][]{das2020ages}. This is shown by the black line where accreted and unevolved stars exhibit lower [Al/Fe] at similar [Mg/Mn] to in situ stars. The last panel shows an alternative projection of [Fe/H] versus [Al/Fe], for which we see the plume of stars reaching high [Al/Fe] at [Fe/H] $\sim$ -1.3 dex for Thamnos (present also in the LE sample) and is likely in situ stars associated to the Aurora population \citep{belokurov2022dawn}. The LE sample is dominated by in situ stars, but does contain a significant accreted or unevolved population.
\textit{Gaia} Enceladus dominates the accreted halo regime and presents a distinct track in [$\alpha$/Fe]-[Fe/H] space along which some Thamnos and Sequoia stars lie.

\subsection{5D Sample}\label{sec:5D}

We now move onto the 5D sample. We used this sample to draw samples with important contributions from Thamnos and Sequoia to characterise their stellar content (see Sect.~\ref{sec:fitting_CMD}).
For our 5D sample, we calculate absolute $G$ magnitude and colour $(G_{\rm BP}-G_{\rm RP})$ correcting for extinction using the \citet{L22} dust maps and using the transformation from E(B-V) to \textit{Gaia} extinction 
from \citet{fitzpatrick2019analysis}.

We then applied several quality cuts on the photometry to obtain clean samples for fitting the CMDs.
We removed stars affected by high extinction, $A_G >0.5$ and  only kept stars with $M_G <$ 5 (the photometric regime for which \textit{Gaia} is complete in our volume of 2.5 kpc) and a good \texttt{phot\_bp\_rp\_excess\_factor} using:
$$
 0.001+0.039 \times \texttt{bp\_rp} < \log(\texttt{phot\_bp\_rp\_excess\_factor})
$$
 and
$$ \log(\texttt{phot\_bp\_rp\_excess\_factor}) < 0.12 + 0.039\times \texttt{bp\_rp}. 
$$
Here, \texttt{bp\_rp} is the apparent $(G_{\rm BP}-G_{\rm RP})$ colour. 
This leaves 30.6 million bright 
5D stars within 2.5 kpc, where the distances are calculated by inverting the corrected parallax, as described in Sect.~\ref{sec:6D_data} for the 6D halo sample. Within this sample, 12.5 million stars have a radial velocity (from \textit{Gaia} or ground-based spectroscopic surveys as described in Sect.~\ref{sec:6D_data}) but the majority of these are thin-disc and thick-disc stars. The CMD of this 5D sample presents stars of all evolutionary stages, including young, bright main sequence stars \citep[as expected from a sample dominated by disc stars e.g.][]{babusiaux2018gaia}. There are 71,977 of the 12.5 million stars with a radial velocity that would be considered halo stars according to |\textbf{V}-\textbf{V}$_\textrm{LSR}$| $>$ 210 km/s. We note that for this sample no quality cuts have been applied to the radial velocity from \textit{Gaia} and no cut in \texttt{RUWE} has been applied since this removes binaries which are included in our synthetic mother CMD (see also Sect.~\ref{sec:dispar}).

\begin{figure}[h]
\centering
\includegraphics[width=0.47\textwidth]{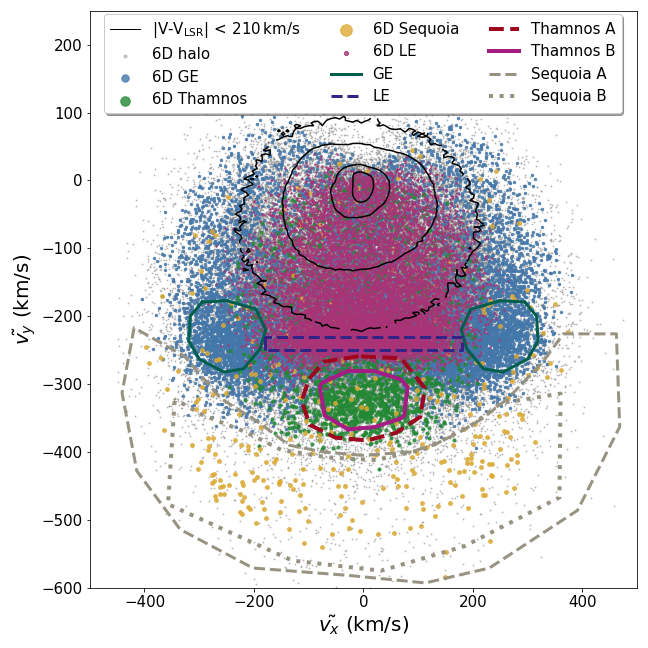}
\caption{Selections of Thamnos, Sequoia, \textit{Gaia} Enceladus (GE), and low-energy (LE) samples in pseudo-Cartesian velocity space, with $\tilde{v}_x$ and $\tilde{v}_y$. Stars with 6D information that belong to the aforementioned structures are shown as coloured markers. The polygons are chosen to select each structure in this space, where the 6D stars pile up, whilst reducing contamination from other structures. The background sample in grey is our full 6D halo sample. Those stars with |$\textbf{V}-\textbf{V}_\textrm{LSR}$| $<$ 210 km/s (i.e. the disc and hot thick disc) are shown with the black contours at levels of 20\%, 68\%, 95\%, and 99\%. 
The LE sample was defined in $v_x-v_y$ space using the same box selection as shown here and  used to select stars that are offset in $v_y$ from Thamnos and in $v_x$ from \textit{Gaia} Enceladus.}

\label{pseudo_selections}
\end{figure}

Although the sample is dominated by disc stars we can quite confidently separate the (accreted) halo stars using 5D information, specifically using the pseudo-Cartesian velocity space, $\tilde{v}_x$, $\tilde{v}_y$, calculated assuming the radial velocity is 0 km/s such that 
\begin{eqnarray}
\tilde{v}_x &=& -v_{l,*}\sin(l) -  v_{b,*}\cos(l)\,\sin(b), \\
\tilde{v}_y &=&v_{l,*}\cos(l) -v_{b,*}\,\sin(l)\,\sin(b), \\
\tilde{v}_z &=&v_{b,*} \cos(b),
\label{eq:1}
\end{eqnarray}
where $v_{i,*} = v_i + v_{i,\odot}$ with $i$ = ($l,b$) and 
\begin{equation}
v_i=4.74057\,\textrm{km}\,\textrm{s}^{-1} 
\bigg(\frac{\mu_i}{{\rm mas~yr}^{-1}}\bigg) 
\bigg(\frac{d}{\textrm{kpc}} \bigg)
\label{eq:2}
\end{equation}
and 
\begin{eqnarray}
v_{l,\odot} & = & -U_{\odot} \sin(l) 
+ (V_{\odot} +V_{\rm LSR}) \cos(l), \\
v_{b,\odot}  &=&W_{\odot}\cos(b) -\sin(b)[U_{\odot}\cos(l)+(V_\odot +V_\textrm{LSR})\sin(l)], 
\label{eq:3}
\end{eqnarray} 
using $ (U, V, W )_\odot$ given in Sect.~\ref{sec:6D_data}.
This is demonstrated in Fig.~\ref{pseudo_selections} which shows where the 6D stars would fall in this pseudo-Cartesian velocity space if we did not know their radial velocity and assumed it to be 0 km/s. The majority of stars with |\textbf{V}-\textbf{V}$_\textrm{LSR}$| $<$ 210 km/s (disc stars) have $\tilde{v}_y \gtrsim -100$ km/s, which does not overlap with the regions of interest in the halo. 

To characterise the Sequoia and Thamnos' stellar populations, we need to be able to select sub-sets of these stars from our 5D sample. As already discussed, we chose to progress to 5D because it is complete (in the range of distance and absolute magnitude relevant to this work) and also allows us to avoid having to model complicated selection functions and, most importantly, it increases the number of stars. However, it is important to note that going to 5D does add additional contamination, which could make the results more difficult to interpret. 

\subsubsection{5D selection of sub-structures}
We followed a two-step process to select the relevant sub-structures in the 5D sample. First, we addressed the pseudo-Cartesian velocity space, $\tilde{v}_x$, $\tilde{v}_y$, calculated assuming the radial velocity is 0 km/s as shown in Eqs. \ref{eq:1}, \ref{eq:2}, and \ref{eq:3}. In Fig.~\ref{pseudo_selections} the 6D Thamnos, Sequoia, GE, and LE samples
are also shown in this pseudo-Cartesian velocity space. The regions where each specific sub-structure dominates and separates from the other structures are outlined by polygons that we used to make a first selection of the structures. 
For Thamnos and Sequoia, we had two selections for each, one of which (B) is more conservative (than A) to minimise contamination at the expense of the number of stars. 
We note that the LE sample selection was defined in the $v_x-v_y$ space using the same box selection as used in pseudo-Cartesian velocity space ($\tilde{v}_x$, $\tilde{v}_y$), which selects a general LE halo sample that is offset in $v_y$ from Thamnos and in $v_x$ from \textit{Gaia} Enceladus.

The second step  involved cleaning each sample further. For this, we followed \citet{ruiz2022_HS_SFH} in assigning each star in the pseudo velocity selections a radial velocity of the closest 6D star (in $l$, $b$, $v_l$, $v_b$) in the same sub-structure; namely, for cleaning the Sequoia sample, we assign to each 5D Sequoia star the radial velocity of the closest true 6D Sequoia star. To select the closest 6D star in $l$, $b$, we used measurement of the distance between two points ($l_1,\,b_1$) and ($l_2,\,b_2$) as $ \cos(\delta) = \sin(b_1)\,\sin(b_2) + \cos(b_1)\,\cos(b_2)\,\cos(l_1-l_2)$ and  for small $\delta$ values, we then assumed $\delta^2 \approx 2\,(1- \cos(\delta)\,)$, scaling this between 0 and 1 our metric for the angular distance on the sky is $d_S^2 = (1- \cos(\delta)\,)/2$. For the closest star in $v_l$, $v_b$, we used the distance metric of $ d_V^2 = (v_{l_1}-v_{l_2})^2/\delta_{v_l}^2 + (v_{b_1}-v_{b_2})^2/\delta_{v_b}^2 $, where $\delta_{v_{l/b}}$ are $\textrm{max}(v_{l/b}) - \textrm{min}(v_{l/b})$ of the 6D sample, such that $ d_V^2 $ is also scaled between 0 and 1 to give equal importance to the distance in velocity space. The total distance to each star is given by $ d_\textrm{tot}^2 =  d_S^2 +  d_V^2 $ and minimising $ d_\textrm{tot}$ gives the nearest neighbour 6D star. 
We then assigned the radial velocity of this nearest 6D star and use this to estimate energy, $E$, and angular momenta ($L_z$, $L_{\perp}$) using the same potential described in Sect.~\ref{sec:6D_data}. 

We selected only those stars whose integrals of motion are within the ellipsoidal selection
of the sub-structure already defined in 6D. 
By definition, any 6D sub-structure stars in the initial pseudo selection will remain in the sample after cleaning in this way but the majority of stars belonging to other sub-structures which overlap in pseudo-velocity space are removed. This method was used before in \citet{ruiz2022_HS_SFH} for the case of the Helmi streams.

\subsubsection{Validation of 5D selected samples}\label{sec:5D_checks}
\begin{table}[h]
    \centering
    \caption{Sizes, completeness, and purity of the samples.}
    \begin{tabular}{|c|c|c|c|c|}
    \hline
         Sample& N in 5D& With $v_\textrm{los}$& Completeness& Purity\\ \hline 
         Thamnos$\,$A & 1773 & 705 & 32\% & 53\%  \\
         Thamnos$\,$B & 1107 & 437 & 20\% & 53\%  \\
         Sequoia$\,$A & 498 & 165  & 41\% & 57\% \\
         Sequoia$\,$B & 406 & 131  & 34\% & 59\% \\
        \hline
    \end{tabular}
    \tablefoot{Samples are selected in 5D and the sizes are given within the fitting bundle (see Sect.~\ref{sec:fitting_CMD}). The purity is only a lower limit, see the text for details. 
    }
    \label{tab:selections}
\end{table}

Table \ref{tab:selections} presents the number of stars in the different sub-sets after first selecting in pseudo velocity space (as shown in Fig.~\ref{pseudo_selections}), followed by cleaning in IoM space as described above and then selecting the stars that are within the CMD fitting bundle (described in Sect.~\ref{sec:fitting_CMD} and shown in Fig.~\ref{fig:CMD_6D}). 
Here, we see the benefit of going to 5D for the smaller sub-structures such as Sequoia, as we have at least three times more stars than if we took the bright 6D samples. 

Along with the contamination already present in the 6D samples, due to the samples not being pure (described in Sect.~\ref{sec:6D_subs}), when we selected our samples in 5D, we had to take into account the fact that this also adds an extra level of contamination due to the missing line-of-sight velocities. In Table \ref{tab:selections}, we present lower limits on the completeness and purity for each sample when compared to the 6D halo in \citet[][extended with spectroscopic surveys; described previously in Sect.~\ref{sec:6D_subs}]{Dodd2023} which we consider to be the `truth'. The completeness for a sub-structure 
(identified in 6D)  was taken to be the ratio of the number of member stars of the sub-structure that fall in the pseudo velocity selection to its total number of (bright) member stars; this is equivalent to the percentage of (bright) member stars of the 6D sub-structure that fall within the pseudo velocity selections in Fig.~\ref{pseudo_selections}. For example, for Sequoia A, the completeness is $N_{\textrm{SeqA \& 6D Seq}}/N_{\textrm{6D Seq \&} M_G<5 }$. 
The purity was calculated as the ratio of 
the number of member stars of the sub-structure that fall in the pseudo velocity selection
to the total number of stars (with 6D information) that enter the pseudo velocity selection (e.g. for Sequoia A, the purity is $N_{\textrm{SeqA \& 6D Seq} }/N_{\textrm{SeqA \& 6D total} }$). 

The purity estimates can be interpreted as lower limits because of our restrictive cut in Mahalanobis distance of 2.13 to assign stars to sub-structures, which was chosen to reduce 
contamination. 
In addition, our 6D samples of the sub-structures, that we are defining as the `truth', are not fully complete themselves. These sub-structures have been defined based on clustering in integrals of motion space and while we were not able to capture the full spread of the merger debris, it can be assumed that this represents the main over-density of the debris. In the local halo sample, only $\sim$ 10\% of the stars are in a significant sub-structure, the remaining will contain accreted stars, including likely stars of GE, Thamnos and Sequoia.

For any given sub-structure, the number of 6D stars that belong to another sub-structure but have still been selected by our criteria using 5D is less than 1\%. However, the 6D stars classified as general halo make up the remaining $\sim$ 40-50\%. This explains why our selection makes the purity appear quite low.
Most of these general halo stars that are added by our 5D method are slightly more extended in IoM space than the cut we used to define our 6D samples (e.g. see Fig.~\ref{fig:5D_check_iom}). However, after an examination of the APOGEE DR17 \citep{accetta2022} abundances of the stars and the metallicity distributions with LAMOST DR7 LRS \citep{zhao2012lamost}, we can conclude that these stars are likely to be true members of the sub-structure (see Appendix \ref{sec:check_5D} for more details) and our 5D selection only adds a small amount of extra contamination.

\section{Methods} \label{sec:method}
\begin{figure*}[h]
\centering
\includegraphics[width=0.95\textwidth]{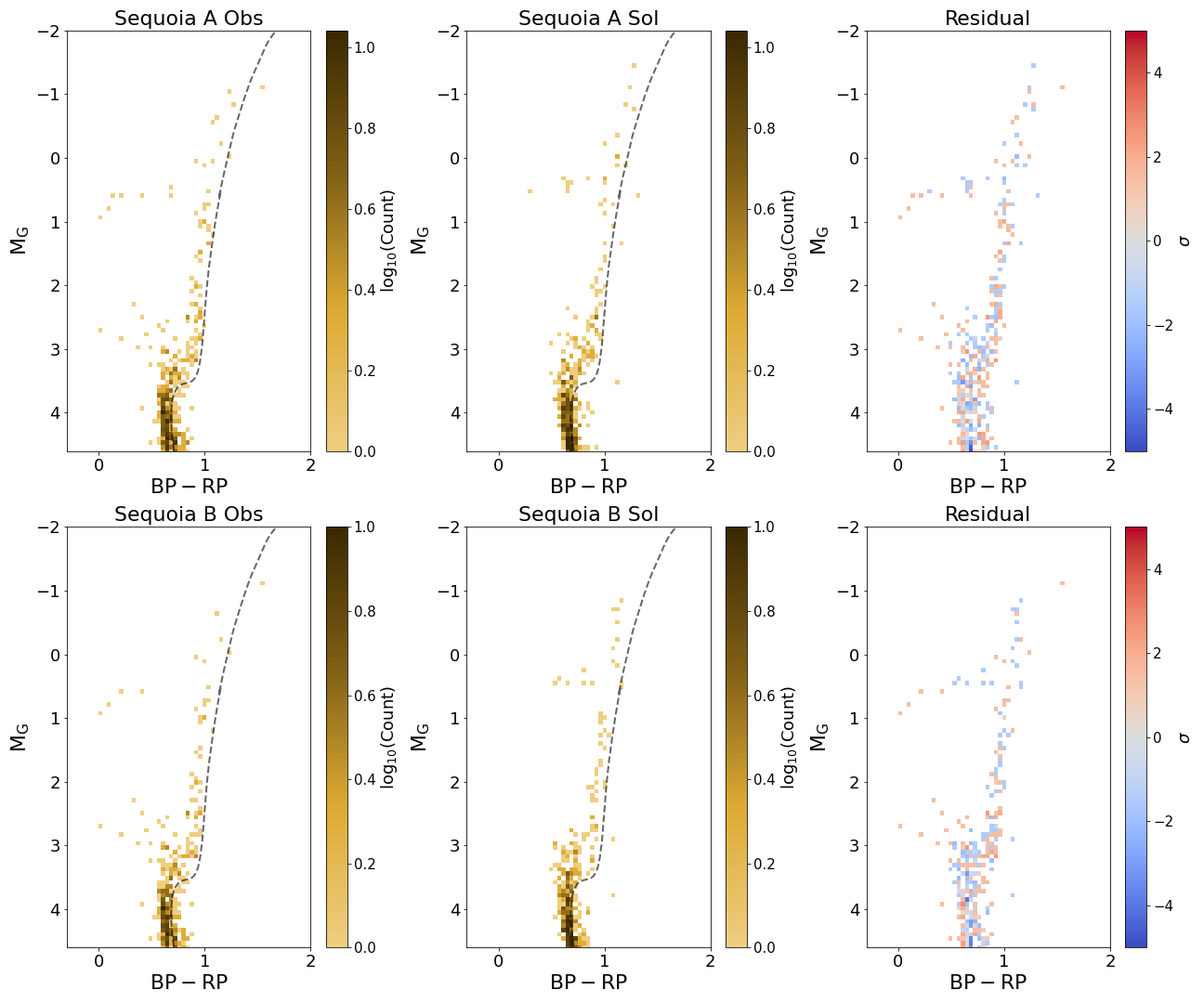}
\caption{Absolute CMDs of the observed 5D Sequoia samples (first column). The top row shows Sequoia A and the bottom row shows the stricter selection Sequoia B. The middle column shows the best fit solution CMD and the final column shows the residual in terms of $\sigma$. The black dashed line in the first two columns corresponds to an isochrone (BaSTI-IAC alpha-enhanced with an age of 11.6 Gyr and [Fe/H] of $-0.8$ dex) that separates the blue and red halo sequences. } 
\label{fig:CMD_seq}
\end{figure*}
\begin{figure*}[h]
\centering
\includegraphics[width=0.95\textwidth]{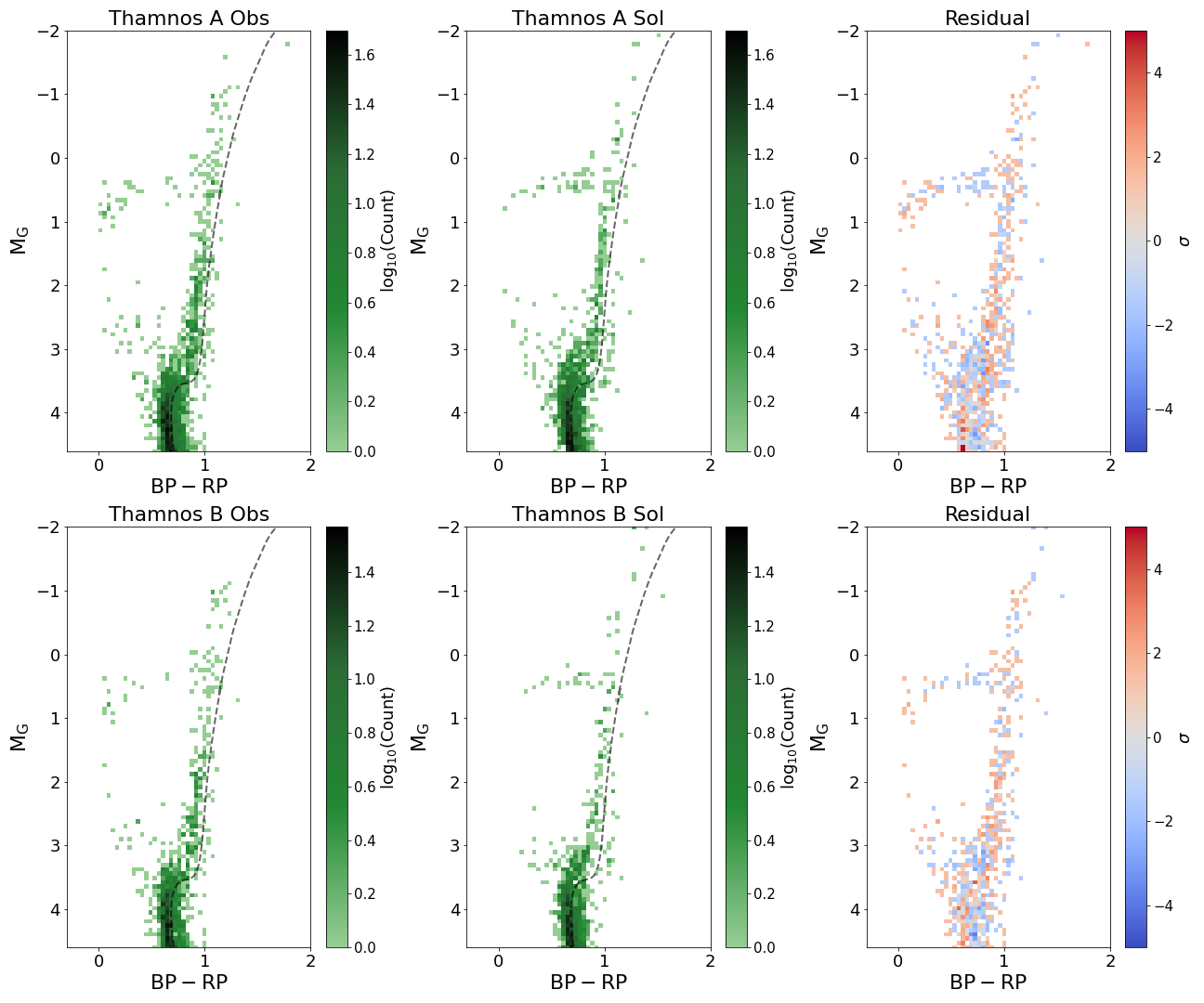}
\caption{Same as Fig.~\ref{fig:CMD_seq}, but for Thamnos. } 
\label{fig:CMDS_Th}
\end{figure*}

By fitting the absolute CMDs of our 5D selected samples, we obtained age and metallicity distributions of their stellar populations. We used an updated method of CMD fitting that has been tailored for use on \textit{Gaia} data, `CMDft.\textit{Gaia}', the details of which are given in \citet[][with an overview given in their Fig. 2]{gallart2024}. This is a three-step process including 1) a synthetic CMD computation (using ChronoSynth, see Sect. \ref{sec:dispar}), 2) then the error and completeness simulation (DisPar-\textit{Gaia},  see \citealt{ruizlara2022}, \citealt{fernandezalvar2025}), and, finally, a search for the best-fit combination of simple stellar populations (SSPs) with DirSFH (a sophisticated update of previous CMD fitting software namely IAC-pop \citealt {aparicio2009iac} and TheStorm \citealt{bernard2018star}). This methodology has also been used to characterise the stellar content of the Helmi streams in \citet{ruiz2022_HS_SFH}.

\subsection{CMD fitting procedure}\label{sec:dispar}
We generated a synthetic mother CMD containing 101 million bright stars ($M_G<$ 5) and a flat distribution in age and metallicity ($Z$) from 0.02 to 13.5 Gyr and 0.0001 to 0.032 (equivalently: [Fe/H] of -2.5 to 0.5 dex) respectively. This CMD was computed using the updated BaSTI-IAC alpha-enhanced models \citep{pietrinferni2021updated} with a Reimers mass loss parameter, $\eta$, of 0.3, a \citet{kroupa1993distribution} initial mass fraction, a minimum mass ratio for binaries of 0.1 and a fraction of unresolved binaries of 30\%.

Our samples are defined within 2.5 kpc of the Sun and contain only stars brighter than $M_G=5$, the magnitude limit for which \textit{Gaia} is >99\% complete in this volume \citep{everall2022completeness}. 
However, in creating our samples we make various quality cuts (e.g. \verb|parallax_over_error| $\geq 5$) which can affect the distribution of stars in the CMD. We therefore must mimic all the same selections in our mother CMD by simulating the quality cuts. To make comparisons between the mother CMD and observations we must also simulate the observational errors. This is done with the step we call DisPar-\textit{Gaia}. For more details of how this is done we refer to Sect. 3.1 of \citet[][]{ruiz2022_HS_SFH} and the comprehensive description presented in Appendix A of \citet{fernandezalvar2025}. This produces an error-convolved mother CMD that is affected by \textit{Gaia} observational errors and selection effects in the same way as the observed sub-samples, allowing for direct comparisons once the possible systematic shift between data and models is also considered. The colour and magnitude shift applied to the mother CMD to match the observations is ($\Delta_c, \Delta_m$) where $\Delta_c$, the shift in ($G_{\rm BP} -G_{\rm RP}$), is taken to be $-0.035$ 
and $ \Delta_m$, the shift in $M_G$, is taken to be $0.04$. The details of how this shift is determined are given in \citet{gallart2024} using the high-quality \textit{Gaia} Catalogue of Nearby stars \citep{smart2021gaia}.

For fitting the observed CMD of our samples using the error-convolved mother CMD we then use DirSFH \citep{gallart2024}. This extracts a series of SSPs from the mother CMD with typical size defined by an array of age and metallicity (Z) seed points
and finds the linear combinations of SSPs which best fit the observed CMD. The spacing of the seed points 
has been chosen to reflect the varying age and metallicity resolutions towards older ages and more metal-poor populations. We perform many realisations of the fit using a Dirichlet tessellation. In each realisation, the seed points are shifted in age-metallicity space. The final solution is the average best fit of all the realisations. 

For this work, we employ the `weighted' strategy within DirSFH which uses the logarithm of the inverse of the variance of ages across the mother CMD to weight the fitting. This allows for more importance to be given to regions of the CMD that have more age resolution \citep[see Sect. 3.3.2 in][]{gallart2024}. Through extensive testing, it was shown by \citet{ruiz2022_HS_SFH} and \citet{gallart2024}  that the solutions remain robust to the choice of weighting, age-metallicity seed points and differences in the mother CMD \citep[see also][as well as Sect. \ref{sec:mocks} of this work for an extensive testing of the method]{gallart2024}.

The fitting of the CMD is done using the stars within a single defined bundle, which can be seen in Fig.~\ref{fig:CMD_6D}. It was defined to encompass all of the stars brighter than M$_G\sim4.5$; namely, chosen to extend $\sim$1 magnitude below the old main sequence turn-off. At magnitudes fainter than this, most of the age information is lost. Since the number of stars increases for fainter magnitudes, they will also dominate in number and, consequently, would be given a lot of importance for fitting, adding more noise than signal, and making the fit less reliable \citep[see Sect. 4.3.2 in][for more details]{gallart2024}. The vertices of this bundle are given in Table \ref{tab:bundle} in Appendix \ref{sec:cmd_fit_parameters}.

\begin{figure*}[h]
\centering
\includegraphics[width=0.95\textwidth]{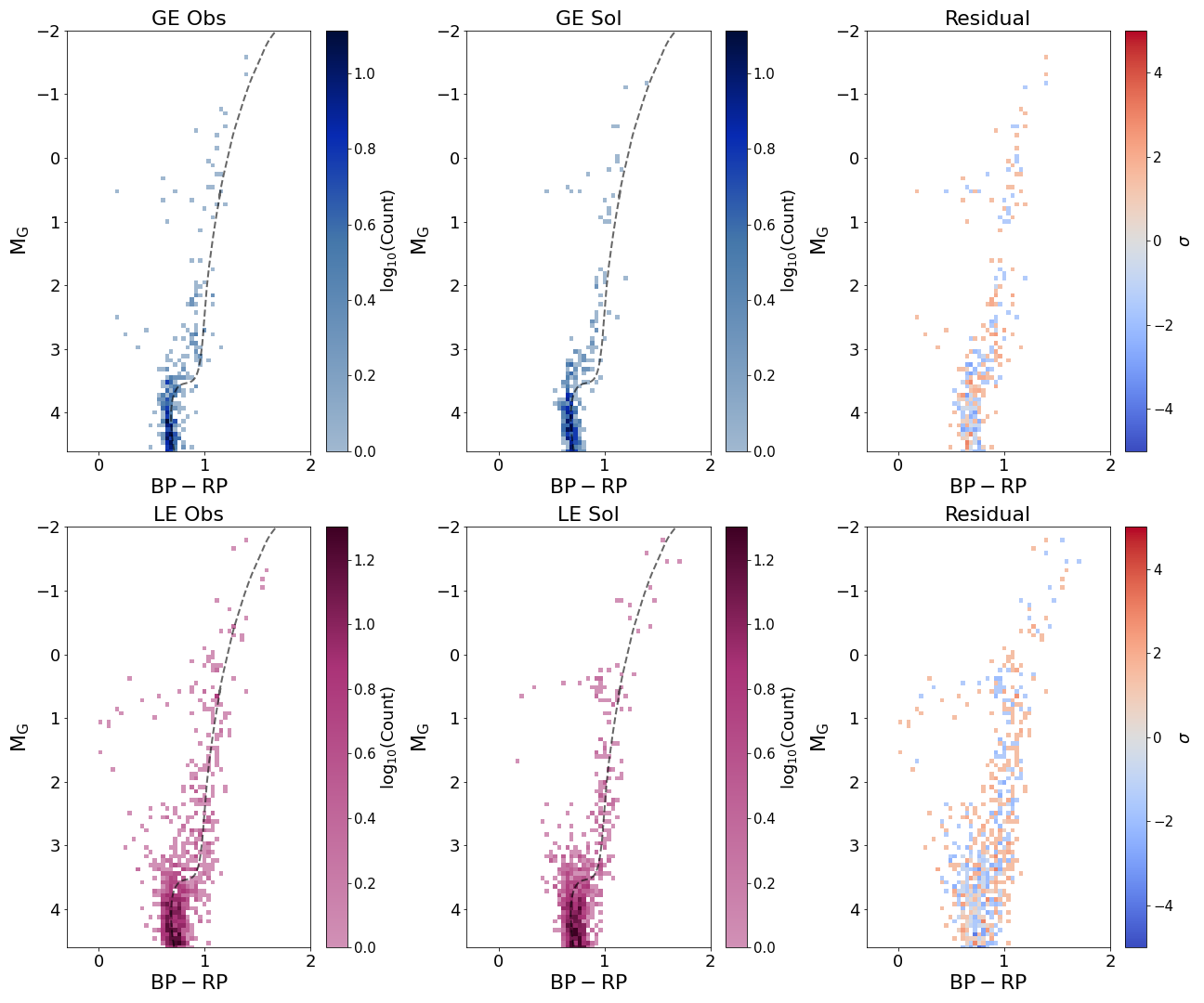}
\caption{Same as Fig.~\ref{fig:CMD_seq}, but for a random sub-sample of the GE and LE samples of sizes $N_S$ and $N_T$, respectively. 
}  
\label{fig:CMDS_GE_LE}
\end{figure*}

\subsection{Fitting the observed CMDs of Sequoia and Thamnos}\label{sec:fitting_CMD}

For each observed sample, we computed 100 solutions of the CMD fitting, such that each time the grid of age-metallicity seeds (see Table \ref{tab:seeds} in Appendix \ref{sec:cmd_fit_parameters} for starting seeds) are slightly modified. DirSFH provides the star formation history (SFH) of the population and the solution CMD (obtained by sampling from the mother CMD according to the best fit SFH, matching the number of stars in the observed CMD within the bundle). Each star in the solution CMD therefore has an associated age and metallicity. Here, we used these best-fit CMDs and the associated age and metallicity distributions to infer differences between the halo sub-structures. We combined the 100 solutions in a weighted average using the goodness of fit as weights. The weighted variance gives our uncertainty on the best fit CMD and on the derived ages, metallicities, and SFHs. 
In our fits, we find that the weights are very similar across all 100 realisations, resulting in the weighted mean and weighted variance being almost identical to the standard mean and variance.

Fitting the CMDs of our samples allows us to obtain age and metallicity distributions for our 5D selections of Sequoia and Thamnos. The CMDs of  Sequoia A and B contain 498 and 406 stars within the bundle respectively. For Thamnos A and B, there are 1773 and 1107 stars within the fitting bundle respectively. The solution CMDs for each sample therefore also contain approximately this number of stars. It was shown in \citet{gallart2024} that this method can achieve very precise ages ($\leq$ 10\%) for single stellar populations, even with $\sim$350 stars, a similar number to our smallest sample. For our samples, we are considering populations with an extended age and metallicity distribution and not a single stellar population so we also check with mock tests that the age and metallicity recovery is precise with this number of stars (see Sect. \ref{sec:mocks} for more details).

In Fig.~\ref{fig:CMD_seq} for Sequoia and Fig.~\ref{fig:CMDS_Th} for Thamnos, the observed and best fit CMDs for each sample are shown, in the left and middle panels respectively. The residuals are shown on the right panels in terms of $\sigma$ =  (obs $-$ sol)$\,$/$\sqrt{(\textrm{obs}+\textrm{sol})/2}$ where obs and sol are the observed and best fit solution CMD respectively. The residuals are small ($< \pm 2 \sigma$) demonstrating that we obtain a good fit for each observed CMD.

We can also validate our 5D samples by comparing their observed CMDs (first column of Figs \ref{fig:CMD_seq} \& \ref{fig:CMDS_Th}) to the 6D samples shown in Fig.~\ref{fig:CMD_6D}, confirming that the 5D are representative of the 6D samples. We see that like in 6D, our 5D samples show that Sequoia and Thamnos exhibit an extended horizontal branch with higher relative contributions of blue HB stars (first column of Figs \ref{fig:CMD_seq} \& \ref{fig:CMDS_Th}) than seen in GE and LE samples (Fig.~\ref{fig:CMDS_GE_LE}). However, this signature is not as well reproduced in our solution CMDs (middle column of Figs \ref{fig:CMD_seq} \& \ref{fig:CMDS_Th}) likely due to low numbers of stars.

HB stars are low mass ($\sim$ 0.5 -- 0.8 $M_\odot$) core helium-burning stars and their presence is indicative of an old stellar population. 
The parameters that determine the morphology of a stellar population's HB have been a long-standing question, with studies of globular clusters (excellent laboratories for stellar evolution) providing evidence (observational and theoretical) suggesting that the main parameter is metallicity, the second main parameter is age and then there is also a strong dependence on the He content \citep[e.g.][]{gratton2010second}. The more metal-poor a stellar population the bluer the HB, the older the population the bluer the HB and the more He enhanced a population the bluer the HB, with the bluest stars forming a tail to lower luminosities. There is not a one-to-one correlation between the HB morphology and the [Fe/H] or age of the population, however, it is still interesting to note that these populations have extended blue HBs
(see Fig.~\ref{fig:CMD_6D}).

\subsection{Fitting the CMDs of contaminants}\label{sec:contaminants}

Both Sequoia and Thamnos samples present contamination from \textit{Gaia} Enceladus debris, and Thamnos also has the additional contamination coming from possibly the in situ halo. As discussed earlier we have represented all of the contamination in Thamnos with our LE sample. Fitting the CMDs of \textit{Gaia} Enceladus and the LE samples will allow us to compare how their contamination presents itself in age-metallicity space.

The 5D selected GE and LE samples contain 4602 (1903 6D) and 6814 (2438 6D) stars respectively.
Since our contaminant populations are significantly larger in the number of stars than Thamnos and Sequoia samples, to be able to compare directly the age-metallicity distributions obtained from CMD fitting, we take random sub-samples matching the number of stars of the relevant sample, within the bundle used for fitting. We choose to match the number of stars within the bundle for the Sequia B and Thamnos B samples, our more restrictive selections with $N_S$ =406 and $N_T$ =1107 stars respectively. That is, 
we take 20 random sub-samples of \textit{Gaia} Enceladus each of size $N_S$ stars and 20 random sub-samples of the LE group each of size $N_T$ stars. As well as having a comparable number of stars for fitting this also allows us to asses any variations in the solutions from sub-sampling. 
The resulting age and metallicity space for our Sequoia and Thamnos samples can now be compared to our \textit{Gaia} Enceladus and LE sub-samples given their similar number of stars and can be used to identify which signatures in age-metallicity space are associated with true Thamnos/Sequoia populations and which are coming from contamination.

Fig.~\ref{fig:CMDS_GE_LE} shows the best fit CMDs with the black dashed line, in the first two columns, corresponding to an isochrone (BaSTI-IAC alpha-enhanced with an age of 11.6 Gyr and [Fe/H] of $-0.8$ dex) that separates the blue and red halo sequences visually. Here we can see that \textit{Gaia} Enceladus populates the blue halo sequence and the LE population contains stars across both sequences but predominantly the red halo sequence, in agreement that this sample is dominated by in situ stars \citep{babusiaux2018gaia}.

\subsection{Mock tests}\label{sec:mocks}
\begin{figure}[h]
\centering
\includegraphics[width=0.5\textwidth]{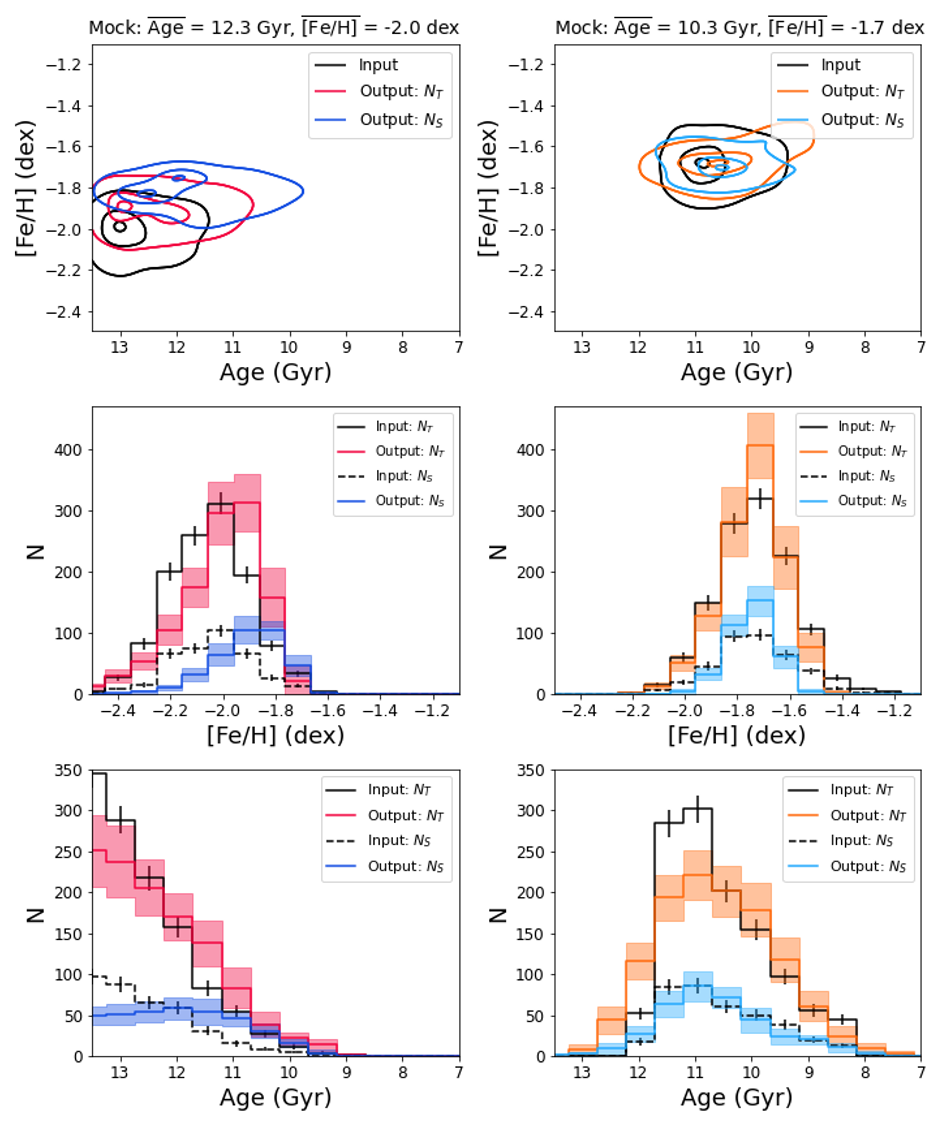}
\caption{Age and metallicity distributions of the input mocks (black) compared to the output of DirSFH. The first column shows the old and metal-poor mock peaking at 12.3 Gyr and -2.0 dex, with blue being the smaller sample ($N_S$ stars) than red ($N_T$ stars). The second column shows the younger, slightly more metal-rich mock (mean age of 10.3 Gyr and [Fe/H] of -1.7 dex) with blue corresponding to the smaller sample of size $N_S$ and orange the larger sample of $N_T$ stars. The top row shows the contours of the 2D distributions containing the percentage of stars within 1, 2 and 3 $\sigma$ levels in 2D. The middle row shows the 1D metallicity distributions and the bottom row the age distributions, where the errors come from averaging over the 100 solutions produced by DirSFH. In the bottom two rows, the input for the samples of size $N_S$ are shown with dashed lines to distinguish between distributions. Errors on the input distributions, plotted as black error bars, are taken as $\sqrt{N}$ per bin.}
\label{fig:toy_mocks}
\end{figure}
To interpret the outputs of the CMD fitting technique, we test the recovery of the age-metallicity distributions of old and metal-poor samples. To create suitable mock populations, we use a mother CMD similar to that used for the CMD fitting as described in Sect. \ref{sec:dispar}. However, since we want to focus on the metal-poor and old regime, we generate more stars here 
 - this mother CMD has an age range of 8 -- 13.5 Gyr, and a metallicity ([Fe/H]) range of -0.5 to -2.5 dex and contains around 30 million stars.

We took sub-samples of this mother CMD following a specific age-metallicity distribution that represents old metal-poor populations like the Sequoia and Thamnos signal. We do this for samples of the same size as Thamnos B and Sequoia B to see how the number of stars affects the recovery of the age and metallicity information. These tests complement those on single stellar populations by \citet{gallart2024} as here we focus on the recovery for extended age and metallicity distributions, concentrating on old ages and low metallicities which is the region of interest for this study.

We selected stars according to a skewed bi-variate normal distribution (eq \ref{eq:mock}), to be able to represent the distributions of ages and metallicities realistically. 
The age distributions are modelled with a high negative skew such that we reproduce the theoretical age distribution for an old accreted population, containing old stars before a decline or cut-off at accretion. The metallicity distributions are not given any skew. 

The form of this distribution is given by:
\begin{equation}
\begin{aligned}
f(\boldsymbol{x} \mid \boldsymbol{\mu}, \boldsymbol{\Sigma}, \alpha_{\textrm{age}}) =   \hskip0.3\textwidth \\
\frac{2}{2\pi \sqrt{|\boldsymbol{\Sigma}|}} \exp \left( -\frac{1}{2} (\mathbf{x} - \boldsymbol{\mu})^\top \boldsymbol{\Sigma}^{-1} (\mathbf{x} - \boldsymbol{\mu}) \right)  \cdot \Phi\left(\alpha_{\textrm{age}} \frac{\textrm{age} - \mu_{\textrm{age}}}{\sqrt{\sigma_{\textrm{age}}^2}}\right)
\end{aligned}
\label{eq:mock}
,\end{equation}
where
$\mathbf{x} = \begin{pmatrix} \textrm{age} \\ [\textrm{Fe/H}] \end{pmatrix}$, 
$\boldsymbol{\mu} = \begin{pmatrix} \overline{\textrm{age}} \\ \overline{[\textrm{Fe/H}]} \end{pmatrix}$, $\boldsymbol{\Sigma} = \begin{pmatrix}
\sigma_{\textrm{age}}^2 & 0 \\
0 & \sigma_{[\textrm{Fe/H}]}^2
\end{pmatrix}$, 
$\Phi(\cdot)$ is the CDF (cumulative distribution function) of the standard normal distribution and
$\alpha_{\textrm{age}}$ represents the skewness in the age distribution.

We make two mocks, one with a $\boldsymbol{\mu}$ of (13.5 Gyr, $-2.0$ dex), and a younger more metal-rich mock with a $\boldsymbol{\mu}$ of (11.5 Gyr, $-1.7$ dex). For both we use a age spread ($\sigma_{\textrm{age}}$) of 2.5 Gyr and metallicity spread ($ \sigma_{[\textrm{Fe/H}]}$) of 0.3 dex. With an age skew $\alpha_{\textrm{age}}$ of -10, this corresponds to age distributions with mean ages of 12.3 Gyr and 10.3 Gyr, respectively. 
We then perform the DisPar-Gaia step to simulate observational errors in the CMDs of our toy mocks and run DirSFH to obtain age and metallicity distributions. 

We can see a comparison of the input and outputs in age and metallicity space in Fig.~\ref{fig:toy_mocks}. We see that for the younger, more metal-rich mock (right column of Fig.~\ref{fig:toy_mocks}) the recovery is better for both sample sizes. For the old and metal-poor mock (left column of Fig.~\ref{fig:toy_mocks}) the peaks of the distributions are shifted to slightly higher metallicities and younger ages. This is especially apparent for the smaller sample size (blue in Fig.~\ref{fig:toy_mocks}), where the shift is more significant. 

These experiments (as well as similar tests not quoted here) are reassuring, although we find that a small sample ($N_S = 406$) is more affected by shifts and an old distribution near the edge of the grid (13.5 Gyr in age) is more affected, making the age distribution less steep and shifting the median to younger ages. 
In all cases, any shift in age and metallicity is always in the direction of younger ages and higher metallicities, with maximum shifts of the mean age on the order of 1 Gyr and of the mean [Fe/H] up to $\sim$0.15 dex. Any effect from being near the edge of the grid that results in a shift to younger ages in the output is always accompanied by a shift to high metallicities as a compensation.

\section{Results}\label{sec:results}

\subsection{Age-metallicity space of best fit CMDs }
We go on to examine the age-metallicity space for our best fit CMD stars of the different samples of Sequoia in equally spaced age bins of 0.75 Gyr and 0.2 dex in metallicity. For these analyses, we used larger bins than the age-metallicity seeds used for fitting the observed CMD, chosen to have enough signal per bin. 

\begin{figure}[h]
\includegraphics[width=0.47\textwidth]{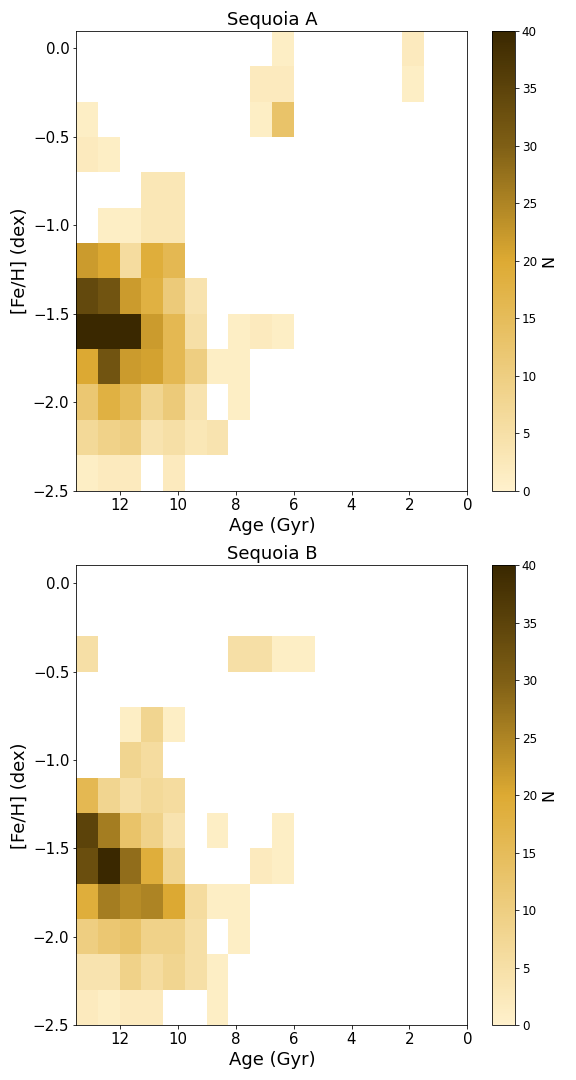}
\caption{Age-metallicity space of best fit CMD stars for Sequoia A and B (with B being the more conservative selection).  }
\label{Seq_age_met}
\end{figure}

\begin{figure}[h]
\includegraphics[width=0.47\textwidth]{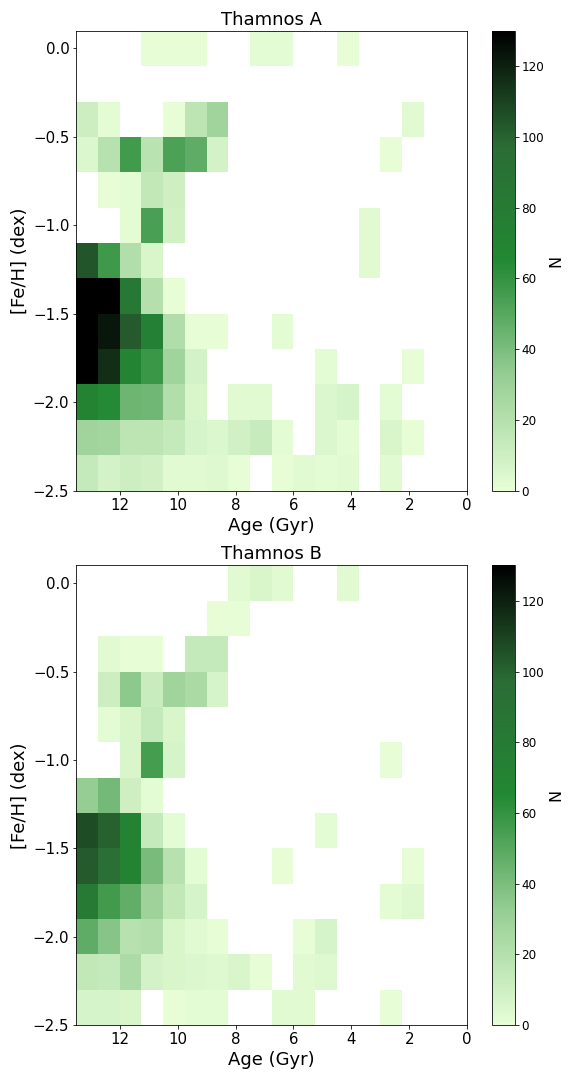}
\caption{Age-metallicity space of best fit CMD stars for Thamnos A and B (B being the more conservative selection).}
\label{Thamnos_age_met}
\end{figure}

Fig.~\ref{Seq_age_met} shows the age-metallicity distribution for Sequoia A
and B. We can see a dominant population of stars around [Fe/H]= $-1.5$ with ages $>$10 Gyr, and a tail to lower metallicities and younger ages. Sequoia A and B (the more restrictive selection) present very similar distributions. 

Fig.~\ref{Thamnos_age_met} shows the age-metallicity space for Thamnos A and B, where Thamnos B is the more conservative selection. In both Thamnos A and B, there is a significant fraction of stars at very old ages ($>$12 Gyr) from [Fe/H] $\sim -1.4$ dex, extending down to [Fe/H] =$-2.0$ dex. We can already see differences 
when comparing to the results of Sequoia in Fig.~\ref{Seq_age_met}; for instance, in our Thamnos selections, there is a higher fraction of stars that are older than $12$ Gyr.

\begin{figure}[h]
\includegraphics[width=0.47\textwidth]{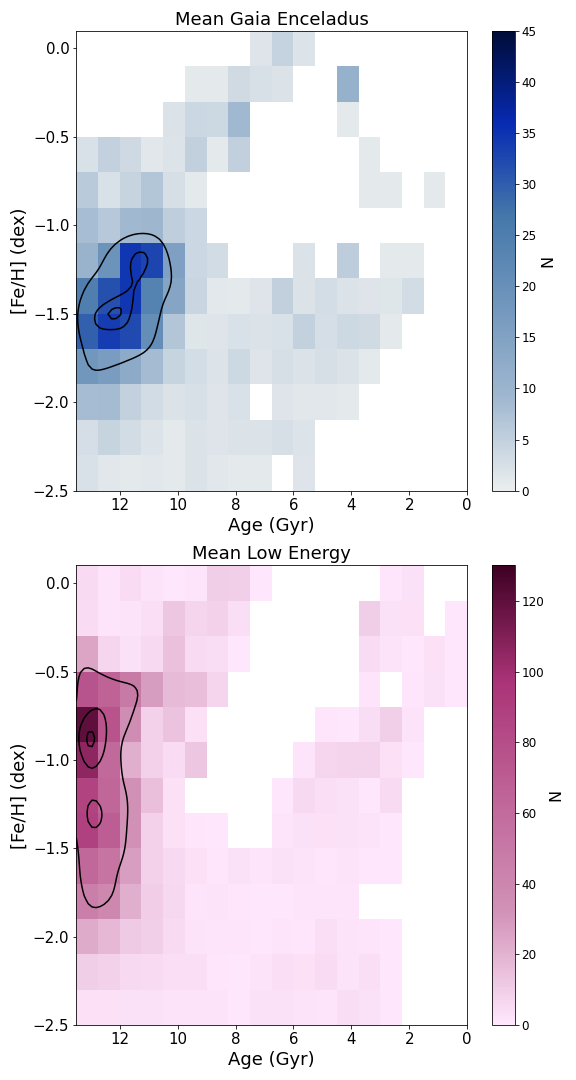}
\caption{Age-metallicity space showing the mean of the best fit CMDs for 20 GE (left) and 20 LE (right) sub-samples. The colour bar shows the count. Contours are shown containing 39\%, 87\% and 99\% of the counts (corresponding to 1, 2, and 3 $\sigma$ in 2D, respectively). }
\label{fig:GE_age_met}
\end{figure}

However, due to contamination, it is difficult to interpret these differences.
This is why we now also characterise in age-metallicity space the contamination that could be present. In Fig.~\ref{fig:GE_age_met} we show the mean age-metallicity distributions from fitting the CMDs of the 20 GE sub-samples (left panel) and that of the 20 LE sub-samples in the right panel. Here, the total number of stars in the LE sub-sample is $N_T$, the same as the number of stars in Thamnos B and in the GE sub-sample is $N_S$, the same as the number of stars in Sequoia B. 
\textit{Gaia} Enceladus presents a well-defined age-metallicity relation extending from [Fe/H] $\sim -2$~dex to $-1$~dex across 13.5 $-$ 10~Gyr in age (see the left panel of Fig.~\ref{fig:GE_age_met}). 
For the LE sample, shown in the right panel of Fig.~\ref{fig:GE_age_met}, we can see that it is dominated by very old stars $> 12$~Gyr, across a wide range of metallicities from [Fe/H] $\sim -1.8$ to $-0.5$ dex, and peaking around [Fe/H]$\sim -0.8$. 
There is a significant tail of stars going down to [Fe/H] =$-2$ dex with ages $> 12$~Gyr old, which could be contributing to the signal in Thamnos.

A visual comparison of the age-met distributions for Sequoia and Thamnos
with those of the expected contamination is already informative.  We
can see that especially Sequoia (and to some extent Thamnos) 
contains a significant fraction of stars that seem to 
follow the \textit{Gaia} Enceladus age-metallicity
sequence. We can also see that Thamnos has a higher
contribution of old stars $> 12$ Gyr, similar to the LE population,
but only at lower metallicities. 
Nonetheless, both Sequoia and Thamnos show a larger fraction of
stars at lower metallicities than in GE and the LE sample. This is the
signal we are interested in. 

\subsection{Comparison of the age-metallicity distributions} \label{sec:compare_2D_distributions}

\begin{figure}[h]
\includegraphics[width=0.47\textwidth]{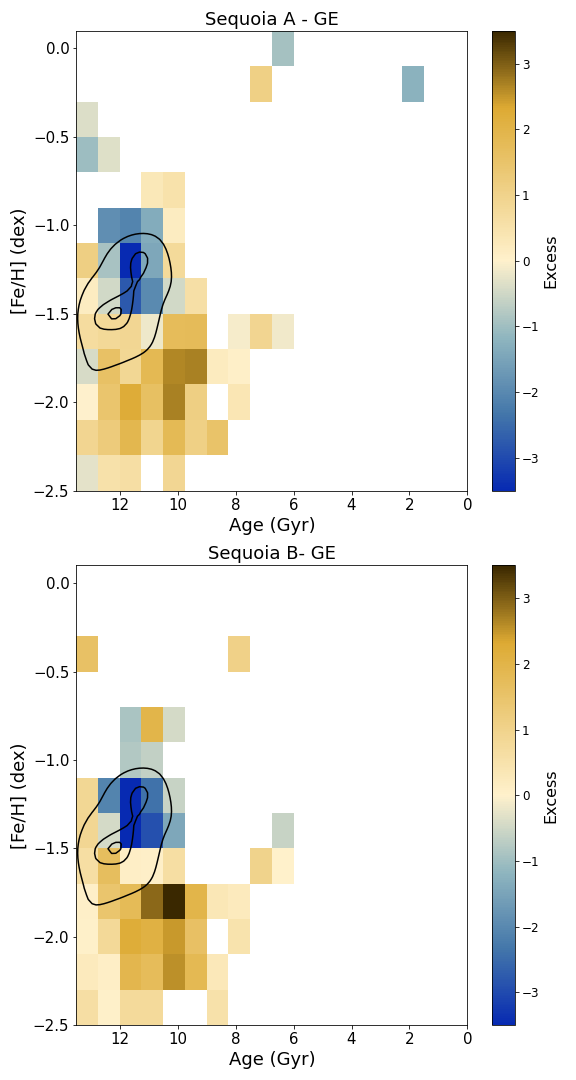}
\caption{Residual age-metallicity space of the GE CMD stars subtracted from those of Sequoia A and B samples (B being the more conservative selection). The colour bar is the difference in the number of stars per bin divided by the combined errors on the count. The contours of the contamination (GE) age-metallicity distribution are shown for visual comparison. }
\label{Seq_age_met_residual}
\end{figure}

\begin{figure}[h]
\includegraphics[width=0.47\textwidth]{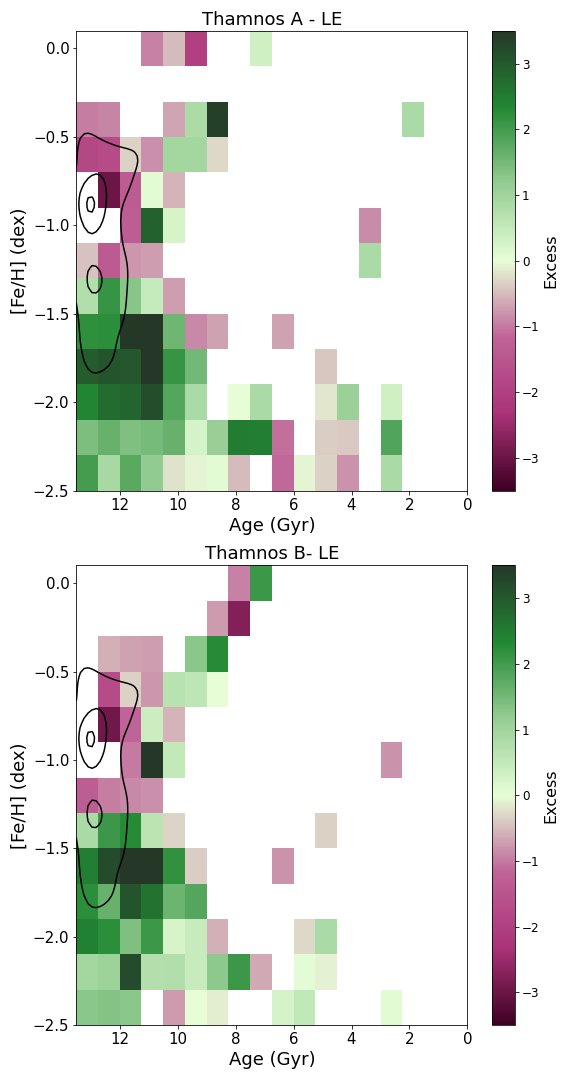}
\caption{Same as Fig.~\ref{Seq_age_met_residual}, but for Thamnos. The residual here is calculated by subtracting the LE sample from Thamnos and the contours are that of the LE age-metallicity distribution, shown for visual comparison.}
\label{Thamnos_age_met_residual}
\end{figure}

To differentiate the signatures coming from
contamination and the signal of Sequoia and Thamnos we subtract the
mean counts in the age-metallicity space of GE from Sequoia
(Fig.~\ref{Seq_age_met_residual}) and of the LE sample from Thamnos
(Fig.~\ref{Thamnos_age_met_residual}). We can subtract our mean
counts of LE from Thamnos~B directly since the samples contain the
same number of stars by definition, $N_T$. The same is true for subtracting the mean counts of GE from Sequoia B, both have $N_S$ stars by definition. For comparison to Sequoia A or Thamnos A which have a different number of stars, we first
introduce a normalisation. In this way, we can directly compare sub-structures and see what fraction of the stars
in the samples dominate in which region of the age-metallicity space. We
scaled the difference in counts with the combined errors to assess the 
impact of variance per bin. For GE and LE, we took the errors,
$\sigma_{GE}$ and $\sigma_{LE}$, to be the standard deviation of
counts in the age-metallicity bins across all 20 sub-samples. This
measures the variation of the solutions and is larger than the Poisson
noise of the mean GE or LE of each sub-sample.

For Sequoia, the residual age-metallicity space after subtracting GE and dividing by the error ($\sigma$ is taken to be $ \sqrt {\sigma_{GE}^2 + \sigma_{S}^2 } $ where $\sigma_{S}$ is the Poisson noise on the counts for Sequoia)
is shown in Fig.~\ref{Seq_age_met_residual}. 
The colour bar gives a measure of the excess of stars in the sample that are not present in the contamination. We see in this figure that there is an excess of stars in Sequoia at low metallicity and up to metallicities of $-1.5$ dex, with ages greater than $\sim$ 9--10 Gyr.

Fig.~\ref{Thamnos_age_met_residual} shows the residual age-metallicity space after subtracting the mean count of the LE sub-samples from Thamnos in each bin and dividing by the error. 
Here the error $\sigma$ is taken to be $ \sqrt {\sigma_{LE}^2+ \sigma_{T}^2 } $ where $\sigma_{T}$ is the Poisson noise on the counts for Thamnos. 
Now we can see that the Thamnos region has an excess of metal-poor stars [Fe/H] $< -1.5$ dex at old ages ($>10$ Gyr) when compared with the contamination.

These figures reveal more clearly that there are subtle differences between Sequoia and Thamnos in their age-metallicity distributions, providing insights into their formation histories and hence possibly on the timing of these mergers. Sequoia reaches similar metallicities as Thamnos ([Fe/H] of $-1.5$ dex) but over a slightly more extended period of time. We will see this more clearly in Sect.~\ref{sec:age_dist}. However, we should bear in mind that this slightly more extended age distribution for Sequoia could be an effect of having a low number 
of stars, which acts in the direction of spreading towards higher metallicities and younger ages, as discussed in Sect.~\ref{sec:mocks}.  
Our findings suggest that Sequoia and Thamnos formed their stars over a similar duration to \textit{Gaia} Enceladus, without enriching to as high metallicities. This can be explained by Sequoia being a smaller progenitor galaxy and Thamnos even smaller still (considering also that Thamnos has a smaller extent in IoM space than Sequoia). 


\subsection{Comparing MDFs of the best-fit CMDs}

\begin{figure}[h]
\centering
\includegraphics[width=0.4
\textwidth]{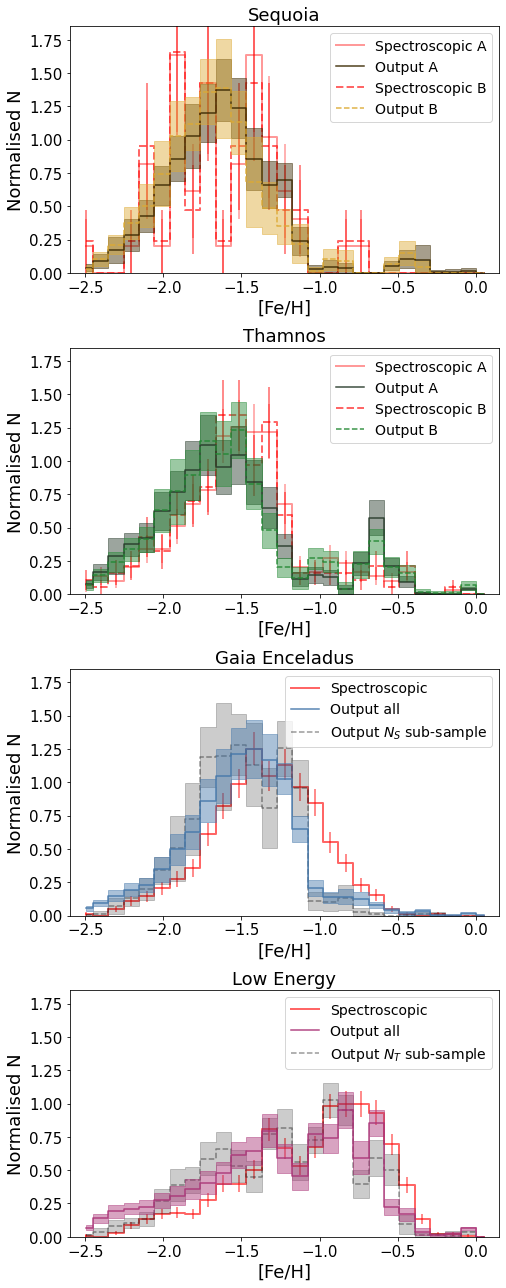}
\caption{Metallicity distributions of the solution CMD stars for the different selections, binned according to the metallicity seeds used for fitting. The solid line shows the weighted mean count per metallicity bin and the shading indicates 1$\sigma$ across 100 iterations. For GE (LE) the output of fitting all stars in the parent sample is shown and one random $N_S$ ($N_T$) sub-sample.
The spectroscopic metallicity distributions (using LAMOST LRS DR7), of the 5D selected samples used for CMD fitting, are also shown here in red. Here, the error bar indicates the Poisson noise ($\sqrt{N}$) in each bin. } 
\label{fig:best_fit_mdfs}
\end{figure}

In Fig.~\ref{fig:best_fit_mdfs}, we compare the metallicity distributions of the solution CMD stars for the different sub-samples. This figure shows the metallicity distributions derived from fitting the CMDs of our samples for Thamnos, Sequoia, GE and LE. The MDFS of Sequoia and Thamnos are obtained by projecting Figs. \ref{Seq_age_met} \& \ref{Thamnos_age_met} onto their y-axis; namely, they still contain the contamination from GE and LE. 
The MDFs presented in Fig.~\ref{fig:best_fit_mdfs} are binned according to the metalicity seeds used for fitting the CMDs, see Sect. 3.2 and Table \ref{tab:seeds}.

We see that the shape of the MDFs derived from fitting the CMDs of our samples closely matches the spectroscopic (LAMOST LRS DR7) MDFs of the samples, which are shown in red in Fig.~\ref{fig:best_fit_mdfs} for comparison.
The photometric MDFs are derived with metallicities coming from BasTI-IAC models and appear to have an offset compared to the spectroscopic metallicities shifting the peaks of the distributions to slightly lower metallicities. We can see visually that there is a shift in the [Fe/H] scale of our photometrically derived metallicities of $\sim$0.1 -- 0.2 dex to lower metallicities. For the Sequoia sample, this is less clear due to the lower number of stars. Despite the offset, the MDFs derived from CMD fitting do recover the same shape and relative distributions as the spectroscopic metallicities.

Sequoia presents an MDF whose peak is shifted to lower metallicities than \textit{Gaia} Enceladus. The same can be said for Thamnos, with some hints at a more prominent metal-poor tail below [Fe/H] $\sim -2$. The LE group presents two main peaks, a large contribution from (hot) thick disc stars with high metallicity  ([Fe/H]$\sim -1.0$ \textrm{to} $-0.5$) and a second more metal-poor peak similar to that of GE debris. It also presents a significant metal-poor tail, as seen in the spectroscopic metallicities, which is the superposition of the metal-poor in situ halo component and metal-poor tail of GE.

\begin{figure*}[h]
\centering
\includegraphics[width=0.75\textwidth]{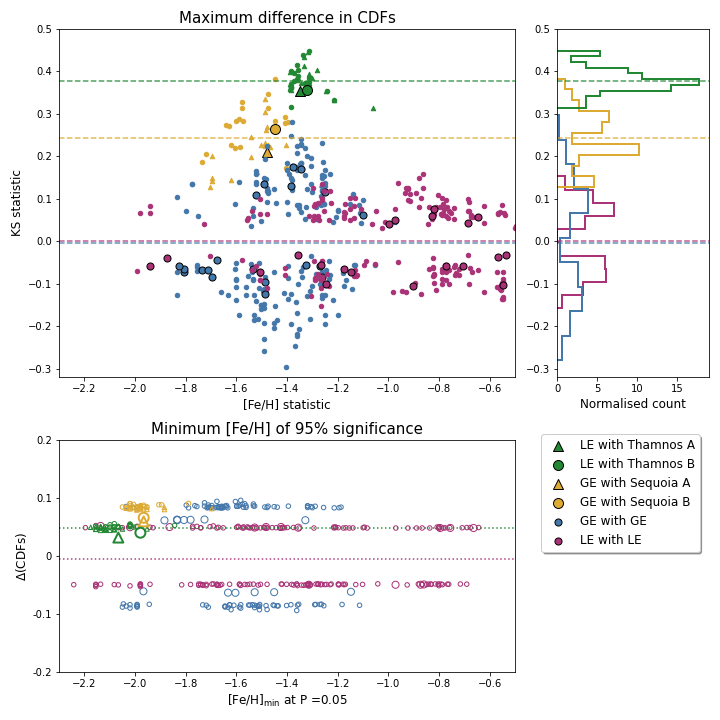}
\caption{Top-left panel: KS statistic (maximum difference) and metallicity of the maximum difference for the comparison of best fit CMD metallicity distributions of various samples. Yellow and green points represent Sequoia and Thamnos samples compared with GE and LE samples, respectively. Large markers represent the comparison of Thamnos/Sequoia A (triangle) with LE and GE and Thamnos and Sequoia B (circle) with LE/GE. Small markers represent the same but with the sub-samples. Blue points show the GE metallicities compared with one another. Pink points show the LE metallicities compared with one another. Large circle markers represent the 20 sub-samples compared with the parent GE/LE and small markers the 20 sub-samples compared with one another. The dashed green (yellow) line represents the mean maximum difference for Thamnos (Sequoia) samples with LE (GE). Top-right panel: Projected distribution of the KS statistics, highlighting the differences between sub-samples and true differences in the case of Sequoia (yellow) and Thamnos (green). Bottom panel: [Fe/H] at which the difference in the cumulative metallicity distributions of various sub-samples becomes significant to the 95\% level. The same colours and markers are used for the different samples. }
\label{KS}
\end{figure*}

To establish  the differences between the MDFs and where
they occur in a quantitative way, we proceeded as follows; we compared the MDFs obtained from
the best fit CMD of GE, including the 20 sub-samples, with one another
and with Sequoia. We did the same for Thamnos compared with the LE
samples. To compare two MDFs, we used the two-sample Kolmogorov–Smirnov
(KS) test. 
Figure~\ref{KS} (top panel) shows
the KS statistic (including the sign of the difference in the distributions) and the [Fe/H] value at which the maximum difference
occurs. Here, the yellow points represent Sequoia A (triangle markers) and
B (circle markers) sample MDFs compared with GE and the 20 GE
sub-samples. Green represents Thamnos A (triangle markers) and B
(circle markers) compared with LE and the 20 LE sub-samples. Blue
points show the GE MDFs compared with one another. Pink points show
the LE MDFs compared with one another. This allows us to quantify how
much variation comes from random sub-samples of GE and LE, and to check that the
differences we measure for Sequoia and Thamnos with the GE and LE samples are significant. The fact that the difference in cumulative metallicity distributions of Sequoia and Thamnos with GE and LE respectively, is always positive tells us that there is an excess in these sub-structures at these metallicities, while the sub-samples compared with themselves have smaller variations with positive and negative sign, which is consistent with being driven by noise.

Looking at the results of the KS tests (shown in Fig.~\ref{KS}) we can see that the largest differences between GE and Sequoia are present at low metallicity. We see a clear difference in the MDFs around [Fe/H] of $-1.5$ to $-1.6$ and this is slightly more significant in Sequoia B which is the stricter sample with lower contamination. 
The mean difference for Sequoia (yellow dashed line in Fig.~\ref{KS}) compared with GE is higher than the differences in the MDFs produced by sub-sampling from \textit{Gaia} Enceladus, meaning that we cannot explain the Sequoia MDF as just a random sub-sampling of \textit{Gaia} Enceladus. 
For Thamnos, in comparison with the LE sub-samples, the maximum difference is of a larger amplitude, since the metallicity distributions become very different as we go to higher metallicities and this maximum difference occurs at [Fe/H] $\sim$ $-1.4$ dex.

\begin{figure*}[h]
\centering
\includegraphics[width=
0.9\textwidth]{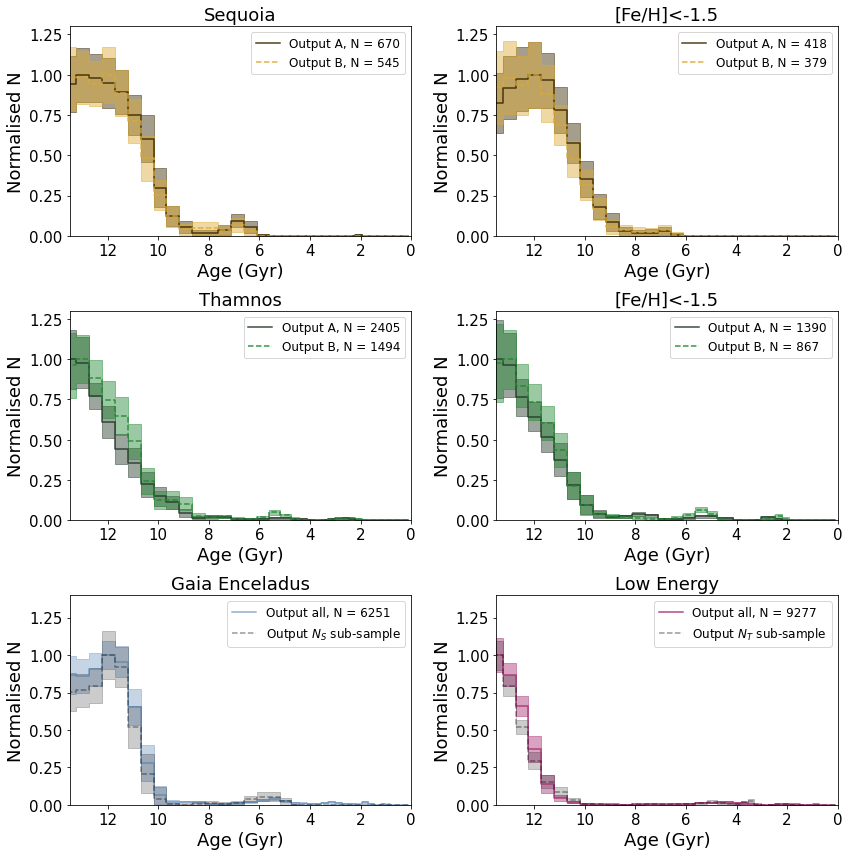}
\caption{Age distributions of the different samples. The line shows the weighted mean count per age bin and the shading indicates 1$\sigma$ across 100 iterations. For the first two rows, Sequoia A (B) and Thamnos A (B) are shown with dashed (solid) lines. The left panels of these first two rows show all stars in the Sequoia and Thamnos best fit CMDs, and the right panels show only stars below a [Fe/H] threshold of -1.5 dex.
The bottom row shows all stars in the GE and LE samples for comparison and one random $N_T$ sub-sample.
}
\label{fig:best_fit_ages}
\end{figure*}

Since the [Fe/H] of the maximum difference in the distributions is not the most informative for detecting the differences in the metal-poor tails, the bottom panel of Fig.~\ref{KS} instead shows  the [Fe/H] at which the sample's cumulative metallicity distributions become significantly different at the 95\% level. Since the value of the difference in the distributions, 
$\Delta$(CDFs), equal to 95\% significance, depends on the sizes of the samples\footnote{The critical value for the two sample KS test of sizes $n_1, n_2$ can be approximated as $c(\alpha) \sqrt{(n_1 + n_2)/n_1\,n_2}$ with 95\% significance corresponding to $c(0.05)$ = 1.36}; then, in the bottom panel of Fig.~\ref{KS}, samples of the same sizes are at a fixed y value. 
For Sequoia, the statistical difference occurs already at [Fe/H] $\sim$ -2.0 dex, and again this signal is always a positive excess in Sequoia. 
For Thamnos, the metal-poor tail of the distribution is even more important, as Thamnos does not contain the metal-rich stars that are present in the LE sample. From Fig.~\ref{KS}, we see that the Thamnos and LE samples are already statistically different at [Fe/H] $\lesssim$ -2.1 dex and this signal is always an excess in Thamnos, confirming the distinct metal-poor tail. The GE/LE samples compared with one another can produce a difference that is significant; however, the sign of the difference is random (consistent with noise) and is typically at higher metallicities. 

We had already seen (in Fig.~\ref{6D_characterisation} and Sect. \ref{sec:6D_subs}) similar differences for the spectroscopic metallicity distributions and now they have been derived for the 5D samples and using our CMD fitting techniques. This provides confidence in the methodology and the soundness of the results obtained.


\subsection{Comparing age distributions of best fit CMDs}\label{sec:age_dist}
Figure~\ref{fig:best_fit_ages} shows the age distributions of the best fit samples. 
For Sequoia and Thamnos in the first two rows of Fig.~\ref{fig:best_fit_ages}, the left panel shows the age distribution for all stars in the sample and the right panel shows stars below a metallicity threshold. This metalicity threshold is chosen to investigate the age distributions of stars where the signal from Thamnos and Sequoia is dominant over contamination; namely, where there is a positive excess in Figs. \ref{Seq_age_met_residual} and \ref{Thamnos_age_met_residual}, and also motivated by where the KS tests show maximum differences with GE and LE (see Fig.~\ref{KS}).

From the full age distributions (see left column of Fig.~\ref{fig:best_fit_ages} ) we can already see that both Sequoia and Thamnos are made up of very old stars (mostly $> 10$ Gyr). Looking at the age distribution below [Fe/H] $= -1.5$ dex (right column of Fig.~\ref{fig:best_fit_ages}) for these sub-structures, we see that for both Sequoia and Thamnos the metallicity cut has little effect on the age distributions. For Thamnos the age distribution is steep with a faster decline than Sequoia.

We also show  the age distribution of all stars in the contaminating samples; GE and LE (bottom row of Fig.~\ref{fig:best_fit_ages}). The first thing to note is that all of Thamnos, Sequoia, and GE have more extended age distributions than the LE population which is very old, containing mostly stars older than 12 Gyr (possibly even older given the edge effects, see Sect. \ref{sec:mocks}). 
The GE sample shows an approximately flat age distribution up to $\sim$ 10 Gyr with a sharp truncation. 
This truncation is different from that seen in Sequoia, which presents a more gradual decline to slightly younger ages than GE. 

For GE, the age-metallicity distribution in 2D (left panel of Fig.~\ref{fig:GE_age_met}) shows two peaks, one at $\sim 12.5$ Gyr and [Fe/H] of $-1.5$, and the other at $\sim$11 Gyr and [Fe/H] $\sim$ -1.3 dex. These two peaks possibly indicate two starbursts in the GE progenitor and could be related to the accretion process (e.g. as the progenitor reaches the first pericentric passage, a burst of star formation could be triggered). We do not see these two bursts in our 1D age distribution, however, since these are the projected ages across all metallicities. 

In a more quantitative sense, looking at the cumulative age distributions, we can estimate the lookback time at which the system formed half of its stars. 
Sequoia A (B) formed half of its stars by a lookback time of $11.9 ^{+0.2}_{-0.3}$ Gyr ($12.0 ^{+0.3}_{-0.3}$ Gyr).
Thamnos A (B) is on average older, does not display a flat age distribution but declines faster, having formed half its stars at $12.4 ^{+0.3}_{-0.3}$ Gyr ($12.3 ^{+0.3}_{-0.3}$).
For both Sequoia and Thamnos samples, we quote here the time at which the cumulative age distributions, with the [Fe/H] cut of $-1.5$ dex applied, reach 0.5, but we find consistent values without the [Fe/H] cut.
For the GE and LE samples, they formed half of their stars by a lookback time $12.1 ^{+0.1}_{-0.1}$ Gyr and $12.9 ^{+0.1}_{-0.1}$ Gyr, respectively. Here, the uncertainties are purely statistical, coming form the 100 different realisations of the CMD fitting and do not take into account possible systematics.

The LE sample age distribution confirms that this population is indeed made up of very ancient stars, in line with the theory that this population contains a significant contribution from the very ancient proto-Milky Way component dubbed Aurora \citep{belokurov2022dawn}. Our age distribution for GE is also consistent with previous studies (including earlier CMD fitting works; \citealt{gallart2019uncovering}, spectra and photometry fitting to MSTO stars; \citealt{bonaca2020timing} and asteroseismology derived ages; \citealt{montalban2021chronologically}) with a truncation at 10 Gyr. Previous studies have found a tail to younger ages \citep[$7-9$ Gyr, e.g][]{horta2024stellar} in GE samples; however, we do not see this here, above the noise.

\section{Discussion}\label{sec:discussion}
We have presented age and metallicity distributions for the retrograde halo sub-structures Sequoia and Thamnos, solely derived  from photometry. This is the first time that age distributions have been provided for these sub-structures \citep[although see][for age distributions of stars in the retrograde halo that they call `Sequoia' which likely contains Sequoia and Thamnos stars with contamination from GE and Aurora]{feuillet2021selecting}. Age information adds an important dimension for understanding the build-up of the stellar halo. Still, we caution that proper handling of contamination is needed for interpreting these small halo sub-structures.

\subsection{Contamination} \label{sec:discussion_contamination}
Throughout, we have highlighted the importance of taking into account contamination from the dominant populations in our local halo (i.e. \textit{Gaia} Enceladus and an in situ halo component) when inferring properties of the progenitors. If contamination is not taken into account we can infer very different progenitor properties (for example, if using the mean metallicity to derive the mass of the progenitor). A more thorough understanding of how \textit{Gaia} Enceladus and the in situ halo are distributed in IoM space will help in interpreting the signal from these smaller accretion events. 
For example, it has been shown in the past that debris from \textit{Gaia} Enceladus are expected \citep[e.g.][]{Koppelman2020,naidu2021reconstructing,amarante2022} and found through chemical abundance studies \citep[e.g.][]{matsuno2022,ceccarelli2024walk} to populate the regions of IoM space where Sequoia and Thamnos are found. The Thamnos region 
is very complex owing to the low-energy orbit and added contamination of an in situ component, dubbed Aurora \citep{belokurov2022dawn}, which is another dominant component at low energy identified with [Al/Fe] abundances. 
While potential stellar population gradients are expected in massive accretion events such as GE \citep[e.g.][]{amarante2022} and present themselves in IoM space, we note that the contamination observed in Sequoia from GE contains stars at [Fe/H] $\sim -1.5$ dex \citep{matsuno2022}. Thus, we do not expect the issue of a metallicity gradient in the progenitor of GE to affect our results, which are based on the low metallicity signal, [Fe/H] $<-1.5$ dex, of Sequoia. 

By selecting samples of \textit{Gaia} Enceladus and the low energy in situ halo that are the contaminating populations, we have been able to compare their age-metallicity distributions to that of Thamnos and Sequoia to unveil the differences. We have shown that there is a clear excess of stars at low metallicity in both the Sequoia and Thamnos sub-structures when compared with the respective contamination (see Figs. \ref{Seq_age_met_residual} and \ref{Thamnos_age_met_residual}).

\subsection{Mocks with contamination}\label{sec:mocks_contamination}
\begin{figure}[h]
\centering
\includegraphics[width=0.5\textwidth]{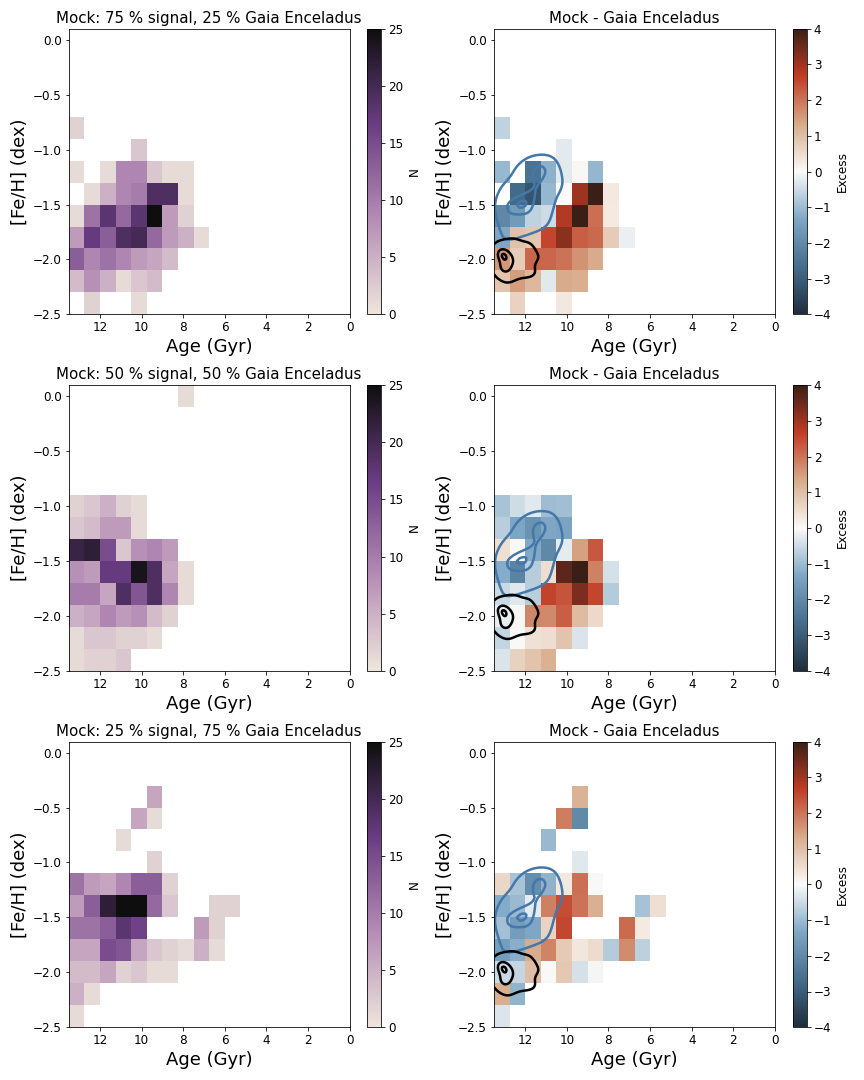}
\caption{Mocks of an old, metal-poor signal with contamination from \textit{Gaia} Enceladus. The top, middle, and bottom rows correspond to the mock with 25\%, 50\%, and 75\% of the stars coming from the old, metal-poor signal and the rest from contamination. In all cases, the total number of stars is $N_S$. The first column shows the outputted age-metallicity distribution from DirSFH and the second column shows this minus the observed \textit{Gaia} Enceladus age-metallicity distribution. The contours show the observed GE distribution (blue) and the black shows the input age-metallicity distribution for the old-metal poor signal.  }
\label{toy_mocks:NS}
\end{figure}
We have previously shown  (Sect. \ref{sec:mocks}) that with the number of stars in our samples, we have the ability to recover an old and metal-poor population. Our resulting [Fe/H] and age distributions resemble the input, but with some shift in the mean to slightly higher metallicities and younger ages. This shift is less than 0.15 dex and 1 Gyr -- always pointing in the direction of slightly younger and more metal-rich stars, where the shift has a slight dependence on how old the population is. 
However, these mock tests were made with a pure old and metal-poor signal and since our observed Sequoia and Thamnos samples also include contamination, we have  constructed mocks that have contamination to check that we can recover an old, metal-poor signal. 

For these new experiments, we took the mocks described in Sect. \ref{sec:mocks} and combined them with sub-samples of the observed contamination samples. 
These mocks consist of an old, metal-poor signal (mean age of 12.3 Gyr and mean [Fe/H] of -2.0 dex) with contamination from \textit{Gaia} Enceladus (Fig.~\ref{toy_mocks:NS}) or the LE sample (Fig.~\ref{toy_mocks:NT}). The mocks are created with varying amounts of the stars (25\%, 50\%, and 75\%) coming from the old, metal-poor signal and the rest from contamination. In all cases, the total number of stars is the same ($N_S$ in Fig.~\ref{toy_mocks:NS}, $N_T$ Fig.~\ref{toy_mocks:NT}). The first column shows the outputted age-metallicity distribution from DirSFH and the second column shows this minus the contamination age-metallicity distribution (mimicking what we show in Figs \ref{Seq_age_met_residual} and \ref{Thamnos_age_met_residual}). 

We see that even for a highly contaminated sample, the old-metal poor signal is recovered in the residual age-metallicity space, but it is always shifted to younger ages and higher metallicities. When the contamination is highest (75\% of the stars, the bottom row of Figs.~\ref{toy_mocks:NS} and \ref{toy_mocks:NT}), then the residual shows an excess population that is distinct from the contamination (at lower metallicities and younger ages); however, this excess does not overlap with the old and metal-poor input sample. The recovered excess is shifted to higher metallicities and younger ages than the input. When contamination is reduced to 50\% and 25\%  
of the total number of stars, then we start to recover signal at the right ages and metallicities as the input signal but the mean is still shifted to younger ages and higher metallicities. This shift is especially prominent for the smaller sample size of $N_S$, as expected from the outcome of toy mocks without contamination in Sect. \ref{sec:mocks}.

Our mock tests do show that for a sample of stars of the same sizes as our observed samples and with realistic contamination, we are still able to recover an excess signal from an extended old and metal-poor population. 
Even though in the residual age and metallicity space, there are shifts compared to the input, we are still able to tell that there is something other than the contamination in these samples. The question of how much this signal is shifted from its true age and metallicity distribution depends on the size of this population and also how it is distributed in age-metallicity space (e.g. if the age-metallicity is correlated, if the stars are predominantly old and at the edge of the grid, how extended the distribution is) and therefore our mocks do not provide an exact quantitative estimate of how the signal will have shifted in our fits of the observed Thamnos and Sequoia samples. However, we note that the residual signal is always shifted to younger ages and higher metallicities, so the signal we find from our observed Thamnos and Sequoia could be older or more metal-poor than what we are finding. Furthermore, the CMD fitting never creates a signal at older ages and lower metallicities than the input population (even when taking into account contamination) and so the observed residuals of Thamnos and Sequoia at lower metallicity (Figs \ref{Seq_age_met_residual} and \ref{Thamnos_age_met_residual}) cannot be created artificially if there is no true old and metal-poor signal present in these samples. 

It has been estimated that the level of contamination from \textit{Gaia} Enceladus in the Sequoia region of IoM space is $\lesssim 20\%$ \citep{matsuno2022}. Our closest mock is shown in the top row of Fig.~\ref{toy_mocks:NS}, consisting of an old and metal-poor sample with contamination from GE making up $25\%$ of the number of stars. The residual signal (top-right panel in Fig.~\ref{toy_mocks:NS}) does qualitatively resemble the signal we see in the observed Sequoia sample in Fig.~\ref{Seq_age_met_residual} with the peak of the signal at ages of $\sim$9--12 Gyr and extending from the lowest metallicities to [Fe/H] $\sim-1.5$ dex (although the observed Sequoia does appear to extend to larger metallicities). This could indicate that the true Sequoia signal is older and more metal-poor than what we observe in Fig.~\ref{Seq_age_met_residual}. However, more tailored tests are needed  
to have a clearer picture of the true Sequoia signal. The fraction of contamination from GE expected in the Sequoia region of IoM space depends on the configuration of the merger and could be a useful constraint in modelling efforts. 

For Thamnos, the highly bound orbit already suggests that contamination is more significant.  Looking at the mock tests in a qualitative sense with contamination from the LE sample (mimicking that expected in Thamnos) as shown in Fig.~\ref{toy_mocks:NT}, the case in which the contamination is $75\%$ is most similar to the observed signal (Fig.~\ref{Thamnos_age_met_residual}). If the contamination is lower than 75\%, then the signal extends to slightly younger ages than those seen in Fig.~\ref{Thamnos_age_met_residual}, with more stars with ages of $<$11 Gyr than what is seen in Thamnos. In these cases, the residual excess shows a strong age-metallicity relationship extending from old ages of $\sim$13 Gyr and [Fe/H] of $-2.5$ dex to ages of $\sim$9 Gyr and a [Fe/H] of $-1.5$ dex. This is not what is seen in what has been observed for Thammos. The observed strength of the excess is qualitatively more similar to the 75\% contamination sample. For lower contamination, the excess signal is stronger and not as representative of Fig.~\ref{Thamnos_age_met_residual}. This suggests that the levels of contamination from general LE populations in the region of Thamnos is of the order of $\sim75\%$ or more, but more tailored tests would be needed.

\subsection{Contamination from 5D selections}\label{sec:disscuss_5D_selection}
We have presented results from CMD fitting of our 5D-selected samples of halo sub-structures. In an ideal case, we would like to make selections in 6D, but we are currently limited by the number of bright stars (oMSTO and above) with radial velocity measurements for the smaller halo sub-structures. We  explain in Sect. \ref{sec:5D_checks} and Appendix \ref{sec:check_5D} that these 5D selected samples resemble the 6D samples in terms of metallicity distributions and chemical abundances. In comparing the CMD of our 5D selected samples (first panel of Figs \ref{fig:CMD_seq}, \ref{fig:CMDS_Th}, and \ref{fig:CMDS_GE_LE}) and 6D samples (Fig.~\ref{fig:CMD_6D}), we have checked that the two agree and confirmed we are not adding different populations to the samples by our 5D selection, although subtle differences are hard to determine. We provided estimates on the completeness and purity of our 5D sample in Sect. \ref{sec:5D_checks}, while both estimates appear quite low, we discuss the fact that our purity is a lower limit and, in fact, the stars that are being added by our 5D selection are generally halo stars that are slightly more extended in IoM space than the sub-structure in question (see Fig.~\ref{fig:5D_check_iom}). These stars that are slightly more extended in IoM space are very likely members of the same population; however, they are not labelled as such due to our conservative selections when defining sub-structures in IoM space \citep{Dodd2023}, since this method is designed to find the main over-densities of debris and sub-structures. 

Both the purity and completeness could still be improved upon by selecting samples in 6D, something that will become possible with \textit{Gaia} DR4 thanks to the increase in radial velocities. However, we reiterate that the 6D selections are not themselves pure and still contain significant amounts of contamination from the general halo (given it is dominated by \textit{Gaia} Enceladus and in situ components). It is not possible to completely remove contamination from the dominant halo components by selecting these small sub-structures purely dynamically, whether in 6D or 5D. The only way to obtain pure selections is to combine chemical selections with dynamic selections of the sub-structures. However, for our CMD fitting, this will not lead to improvements unless the selection function and completeness effects are very well understood and can be properly modelled.

\subsection{Selection of Thamnos}\label{sec:Th_selection}
\begin{figure}[h]
\centering
\includegraphics[width=0.36
\textwidth]{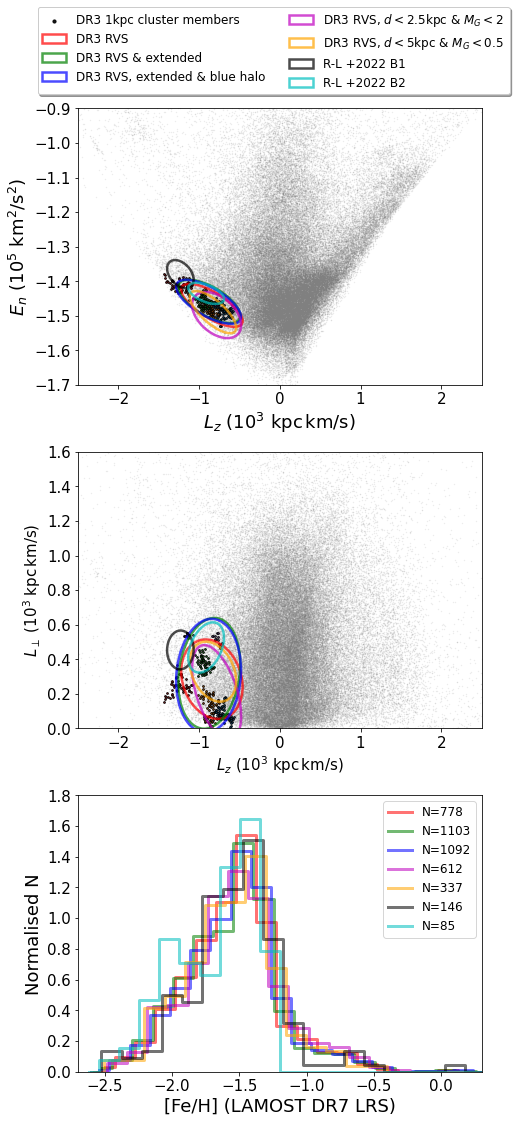}
\caption{IoM space of the local halo sample described in Sect. \ref{sec:6D_data} with ellipsoids showing over-density definitions of Thamnos from running a clustering algorithm \citep{lovdal2022} on various samples. The \textit{Gaia} RVS sample definition used in this work is shown by a red ellipsoid. Other selections include extending the RVS sample (addition of ground-based RVs before clustering instead of adding them to the pre-defined clusters), a blue halo selection (those that are left of the isochrone separating the two sequences shown in e.g. Fig.~\ref{fig:CMD_seq}), and complete samples (taking the $M_G$ at the level that all stars down to that magnitude, in that volume, have a $g_{RVS}$ above the RVS limit). The definitions by \citet{ruizlara2022} of Thamnos 1 and 2 are shown as black and cyan ellipsoids here (see R-L+2022 in the legend). Black points show the individual clusters in the Thamnos region when clustering on a 1kpc halo sample, highlighting the sub-structure we observe. The bottom panel shows the LAMOST DR7 LRS metallicity distributions for selecting from our 6D halo sample (Sect. \ref{sec:6D_data}) according to the different ellipsoids. These are normalised so that the area of each is equal to 1 and binned each with a relative shift of 0.01 dex to aid the comparison.
}
\label{fig:th_selections}
\end{figure}

Our selection of sub-structures was motivated by the results of clustering in IoM space on \textit{Gaia}DR3 data presented in \citet{Dodd2023}. Prior to this, the same algorithm was applied to \textit{Gaia} EDR3 data (the main difference between the two data releases is a factor of 4.7 less radial velocities in \textit{Gaia} EDR3) supplemented with ground-based spectroscopic survey radial velocities \citep{lovdal2022,ruizlara2022}. We found limited differences between the two datasets with regard to the definitions of sub-structures that are relevant to this work. However, one difference worth noting is the definition of Thamnos. The definition from \citet{ruizlara2022} of Thamnos is shown in Fig.~\ref{fig:th_selections} as black and cyan ellipsoids in IoM space. Here Thamnos is split into two sub-structures; Thamnos 1 at higher energy and Thamnos 2 at slightly lower $L_z$. Compared to our current DR3 definition (red ellipsoid in Fig.  \ref{fig:th_selections} and green points in Fig.~\ref{6D_characterisation}) we can see that the ellipsoidal contours of \citet{ruizlara2022} are smaller and at the higher $L_\perp$ end of our distribution, with Thamnos 1 being at higher energy than our definition. We find that in our clustering run on DR3, if we lower the significance threshold for detecting a cluster, we start to have sub-structures overlapping with Thamnos 1 in $E_n$. 

The other ellipsoids shown in Fig.~\ref{fig:th_selections} correspond to running the clustering algorithm on different samples and are shown here to highlight the robustness of the Thamnos selection in this IoM region. These samples include extending the \textit{Gaia} RVS sample with ground-based spectroscopic survey RVs before clustering (green ellipsoid in Fig.~\ref{fig:th_selections}) and combining this with a blue halo selection (blue ellipsoid in Fig.~\ref{fig:th_selections}) by selecting stars bluewards of an isochrone that separates the two sequences (BaSTI-IAC alpha-enhanced with an age of 11.6 Gyr and [Fe/H] of $-0.8$ dex; see dashed line in the first two columns of Figs \ref{fig:CMD_seq}, \ref{fig:CMDS_Th}, and \ref{fig:CMDS_GE_LE}). Since a radially incomplete sample can produce artificial overdensities in IoM space \citep[and specifically along parabolic lines in $L_z-E_n$ space, see][]{lane2022kinematic} we also run our clustering algorithm on three other samples: 1\,kpc in which we are complete in \textit{Gaia} RVS, 2.\,5kpc with a cut of $M_G < 2$ such that all stars down to that magnitude will have a \textit{Gaia} RV in that volume and 5\,kpc with a cut of $M_G < 0.5,$ such that all stars in this volume will have a \textit{Gaia} RV. 
The individual clusters in the Thamnos region within a 1\,kpc sample are shown in Fig.~\ref{fig:th_selections}, highlighting the individual sub-structure seen in this region. 
Despite these different samples, we can see that our definition of Thamnos remains very similar and we see that also in the metallicity distributions of the selected stars (see bottom panel of Fig.~\ref{fig:th_selections}), there is always a peak around [Fe/H] $\sim -1.5$ dex and then a significant metal-poor tail. 

One reason we highlight the difference between \citet{ruizlara2022} and our Thamnos definition here is that Thamnos 1 in \citet{ruizlara2022} has a larger contribution of metal-poor stars and a more significant metal-poor tail (cyan histogram in the bottom panel of Fig.~\ref{fig:th_selections}), likely due to lower contamination from the in situ halo at higher $E_n$ and $L_\perp$ and it being more retrograde. This was  already seen in the discovery of Thamnos 1 and 2 in \citet{koppelman2019multiple}; namely, in their Fig. 2 with $Lz-E$ colour coded by [Fe/H] we see an excess of very metal-poor stars ([Fe/H]$\sim-2$ dex) at the higher $E_n$ region of Thamnos. The number of stars here is very low and likely similar to the number of metal-poor stars we are seeing at slightly lower $E_n$ in our sample, but the relative contamination is greatly reduced so it appears to be a purer selection. If we were to select a Thamnos 1 sample from our \textit{Gaia}DR3 halo sample (supplemented with ground-based RVs; see Sect. \ref{sec:6D_data}), according to the definition in \citet{ruizlara2022}, then we would only have  209 stars (172 with $M_G < 5$). This was not a sufficient number of stars for the CMD fitting, which is why we stuck to our DR3 definition and modelled the contamination carefully.

\begin{figure}[h]
\centering
\includegraphics[width=0.36
\textwidth]{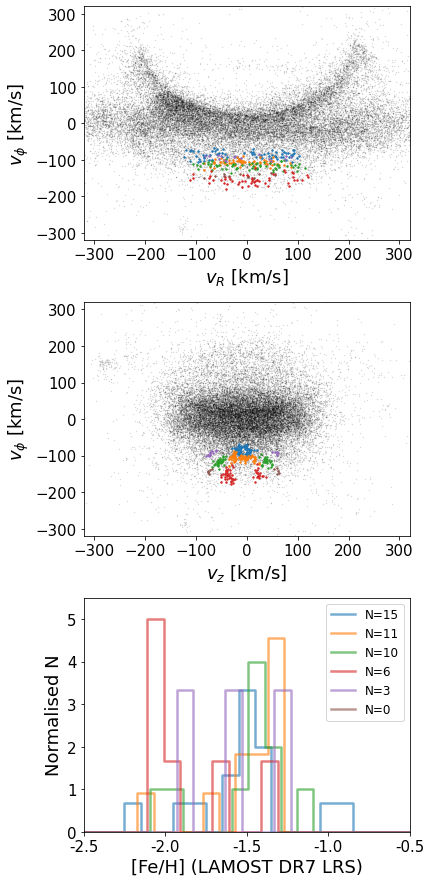}
\caption{Velocity space of local 1kpc halo sample with significant clusters in the Thamnos region highlighted in different colours. We can see the sub-structure presents itself as tight clumps in $v_\phi$ and $v_z$. The bottom panel shows the LAMOST DR7 LRS metallicities (normalised such that the area sums to 1 and binned with a relative shift of 0.01 dex to aid comparison), indicating that some of these velocity sub-structures could be tight in terms of metallicity, but they always present some contamination from the [Fe/H] $\sim -1.5$ dex peak. 
}
\label{fig:1kpc_vel}
\end{figure}

Thamnos occupies a region of IoM space that is very difficult to disentangle due to the low-energy orbit and the expected overlap of debris in this region, including with the tails of other larger sub-structures (\textit{Gaia} Enceladus and Aurora) as discussed throughout this paper. However, the complexity becomes even greater when we look at the Thamnos region in more detail. There is a large amount of sub-structure within this region that can be picked up by clustering in IoM space. In Fig.~\ref{fig:th_selections}, we show the results of clustering on the \textit{Gaia} RVS halo sample confined to a 1\,kpc volume around the Sun, the significant ($>3\sigma$) clusters are shown in black points. Looking at the velocity sub-structure in this small volume of 1kpc around the Sun (Fig.~\ref{fig:1kpc_vel}), we can see these clusters correspond to distinct and tight velocity sub-structures. An accreted sub-structure within a small volume is expected to display clumping in velocity space corresponding to individual streams on different orbital phases. Likely we are seeing the individual streams within the Thamnos debris. Some of these individual streams are very tight in velocity and with limited metallicity information (from LAMOST DR7 LRS: see bottom panel of Fig.~\ref{fig:1kpc_vel}) show a small spread (e.g. red points in Fig.~\ref{fig:1kpc_vel} show a small spread with four stars around [Fe/H] of $-2.0$ dex, one at [Fe/H] $-1.6$ and $-1.3$ dex). It would be interesting to follow these sub-clumps up further.

In summary, the accreted debris, which we refer to as Thamnos, appears to be present across quite an extended range of $L_z,\, L_\perp \& \, E_n$ (see Fig.~\ref{fig:th_selections}), as we see an excess in metal-poor stars. 
A more detailed look at this region with high-precision chemical abundances is needed to provide more insights into Thamnos' debris.
It is also important to note that if we make the in situ versus accreted selections suggested by \citet{belokurov2022dawn} we lose the Thamnos sub-structure, which we have shown here is distinct from the general low-energy stellar populations. 

\section{Conclusions}\label{sec:conclusion}
In this paper, we  present, for the first time, age distributions of the retrograde halo sub-structures: Sequoia and Thamnos. These data have been derived purely photometrically using CMD fitting techniques \citep[i.e. CMDft.\textit{Gaia}:][]{gallart2024} applied to samples of stars extracted from \textit{Gaia} 5D data (missing radial velocities). CMD fitting also provides us with metallicity distributions that successfully reproduce the spectroscopic metallicity distributions of these samples (with a mean offset of $\sim$ 0.1--0.2 dex to lower metallicities; see Fig.~\ref{fig:best_fit_mdfs}) testing the capability of CMDft.\textit{Gaia} on old, metal-poor, and extended populations. We have shown that these age and metallicity distributions are well recovered, even with small sample sizes on the order of $\sim$ 400 stars and with extended age-metallicity distributions as expected for old accreted debris.

Our analysis has shown that Sequoia and Thamnos present metal-poor populations that are statistically different from the contamination expected in those regions of IoM space. From Figs \ref{Seq_age_met_residual} and \ref{Thamnos_age_met_residual}, we can see that both present an excess of stars below [Fe/H] $\sim$ -1.5 dex. For Thamnos, these stars are mostly older than 11 Gyr; howver, for Sequoia, this extends to slightly younger ages of $\sim$9 Gyr at [Fe/H] of -1.5 dex. However, we caution that with the lower number of stars in our Sequoia samples, we could expect some shifts in the distribution to younger ages and higher metallicities. This is shown by our mock tests (see Sects. \ref{sec:mocks} \& \ref{sec:mocks_contamination}), which could explain this signal and the shape of the age-metallicity distribution seen in Fig.~\ref{Seq_age_met_residual}. 

We have looked at the projected age distributions of Thamnos and Sequoia stars below [Fe/H] of -1.5 dex to compare the differences with the full population (dominated by contamination) in Fig.~\ref{fig:best_fit_ages}. The metallicity cut shows a distribution that remains close to constant up to $\sim$11 Gyr for Sequoia, before a gradual decline to $\sim$8 Gyr. This is different from what is seen for \textit{Gaia} Enceladus, where the decline (signifying the accretion) is a lot faster. We are not able to say whether this slower decline in Sequoia is significant or a limitation of the number of stars in our CMD fitting, but it could be telling us about the process and duration of accretion for the Sequoia progenitor. It would be interesting to look into this further with a larger sample of stars. 

Thamnos is strikingly different in its age distribution showing a decline already from $>$13 Gyr ago. This could mean a very early accretion, earlier than \textit{Gaia} Enceladus, which would be in line with the low-energy orbit. 
Since there is still an important contribution from the contamination at low energy (see e.g. the LE contours in Fig.~\ref{Thamnos_age_met_residual} compared to residual Thamnos signal), we have checked that below [Fe/H] $< -2.0$ dex, we still see this steep decline in the Thamnos age distribution. Thus, the result appears to be robust.

The in situ low-energy (LE) population is dominated by the oldest stars in the local halo, telling us that these stars had already formed before the accretion of \textit{Gaia} Enceladus, Sequoia, and Thamnos.
For \textit{Gaia}-Enceladus, we find that CMDft.\textit{Gaia} is now able to uncover an age-metallicity relation (Fig.~\ref{fig:GE_age_met}) and this can be investigated in more detail with 6D selections due to the large number of stars in this sub-structure.

The fact that the age distributions of \textit{Gaia} Enceladus match predictions by other studies \citep[e.g.][ using different methods and data]{bonaca2020timing,montalban2021chronologically} and that the findings for the timeline for accretion of Sequoia and Thamnos match predictions based on their orbits (earlier accretion events sink to lower energy orbits) highlights the power of CMD fitting for uncovering the chronology of the assembly of our halo. This finding, combined with the fact that the spectroscopic metallicity distributions match our photometric metallicities very well, allows us to have substantial confidence in the results provided by CMDft.\textit{Gaia} \citep{gallart2024}. This will serve as an important tool for providing the age information that we have been so desperately missing. We show that this tool can already begin to be used, in combination with the rich \textit{Gaia} dataset, to map out the local halo providing the temporal axis that is enabling us to interpret halo sub-structures and piece together the build-up of our halo. 

With the upcoming \textit{Gaia} DR4, we will have access to  more stars with radial velocity information, which  will allow us to increase our sample sizes for more reliable determinations of the age and metallicity distributions. We can combine this with detailed chemical abundance information to obtain a multi-dimensional understanding of our local halo. 
An interesting avenue would be to map the low-energy regime in terms of ages and understand how these ambiguous populations change over their evolution.  Another interesting comparison is to look at the globular clusters (GCs) associated with these different sub-structures and see where they lie in age-metallicity space to confirm their association. This is the main aim of the CARMA project \citep{massari2023cluster}, namely,  it has been designed to provide a very accurate chronological map of the Galactic GCs system that is fully consistent with that provided by CMDft.\textit{Gaia} and  based on the same stellar evolutionary framework (i.e. BaSTI-IAC).

\begin{acknowledgements}
The authors thank the anonymous referee for their helpful and constructive feedback. This research has been partially funded by a Spinoza award by NWO (SPI 78-411) and support from a MW-Gaia COST (European Cooperation in Science and Technology) action (CA18104) STSM (Short-Term Scientific Mission) grant. 
TRL acknowledges support from Juan de la Cierva fellowship (IJC2020-043742-I) as well as research projects PID2020-113689GB-I00, and PID2020-114414GB-I00, financed by MCIN/AEI/10.13039/501100011033, the project A-FQM-510-UGR20 financed from FEDER/Junta de Andaluc\'ia-Consejer\'ia de Transformaci\'on Econ\'omica, Industria, Conocimiento y Universidades and by grants P20-00334 and FQM108, financed by the Junta de Andaluc\'ia (Spain). 
EFA acknowledges support from HORIZON TMA MSCA Postdoctoral Fellowships Project TEMPOS, number 101066193, call HORIZON-MSCA-2021-PF-01, by the European Research Executive Agency. 
CG, TLR, SC and EFA acknowledges support from the Agencia Estatal de Investigación del Ministerio de Ciencia e Innovación (AEI-MCINN) under grant “At the forefront of Galactic Archaeology: evolution of the luminous and dark matter components of the Milky Way and Local Group dwarf galaxies in the \textit{Gaia}era” with reference PID2020-118778GB-I00/10.13039/501100011033. 
CG also acknowledges financial support from the Spanish Ministry of Science, Innovation and University (MICIN) through the 
Spanish State Research Agency, under Severo Ochoa Centres of Excellence Programme 2020-2023 (CEX2019-000920-S). 
SC acknowledges financial support from PRIN-MIUR-22: CHRONOS: adjusting the clock(s)  to unveil the CHRONO-chemo-dynamical Structure of the Galaxy (PI: S. Cassisi) finanziato dall’Unione Europea – Next Generation EU, and Theory grant INAF 2023 (PI: S. Cassisi).
This work has made use of data from the European Space Agency (ESA) mission
{\it Gaia} (\url{https://www.cosmos.esa.int/gaia}), processed by the {\it Gaia}
Data Processing and Analysis Consortium (DPAC,
\url{https://www.cosmos.esa.int/web/gaia/dpac/consortium}). Funding for the DPAC
has been provided by national institutions, in particular, the institutions
participating in the {\it Gaia} Multilateral Agreement.
The analysis has benefited from the use of the following packages: vaex \citep{breddels2018}, numpy \citep{van2011}, matplotlib \citep{hunter2007}, scipy \citep{2020SciPy-NMeth}, seaborn \citep{Waskom2021}, swiftascmaps \citep{borrow_2021_5649259} and jupyter notebooks \citep{kluyver2016}.

\end{acknowledgements}

\bibliographystyle{aa} 
\bibliography{references}

\begin{appendix}

\section{Validity of our 5D selections}\label{sec:check_5D}
\begin{figure}[ht]
\centering
\includegraphics[width=
0.3\textwidth]{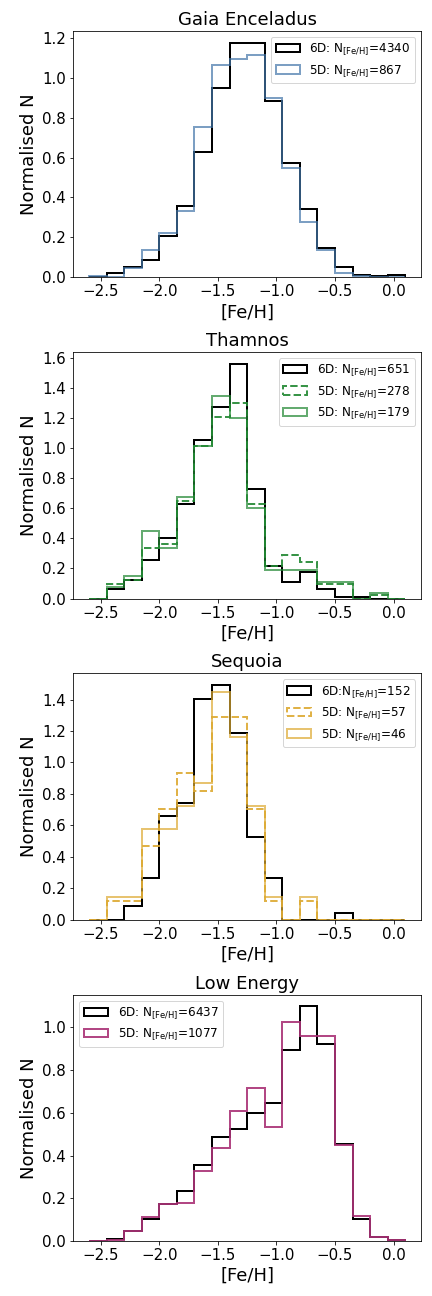}
\caption{LAMOST DR7 LRS metallicity distributions of the 6D sub-structures and the 5D selected samples. Since the 5D method removes some 6D stars (those outside of the pseudo velocity selection) and adds stars, we can check the metallicity distributions of the stars that have a LAMOST DR7 LRS [Fe/H] that have been added to check if we are adding noise or if the population resembles the original 6D defined population.}
\label{fig:5D_check}
\end{figure}

\begin{figure*}[ht]
\centering
\includegraphics[width=
0.8\textwidth]{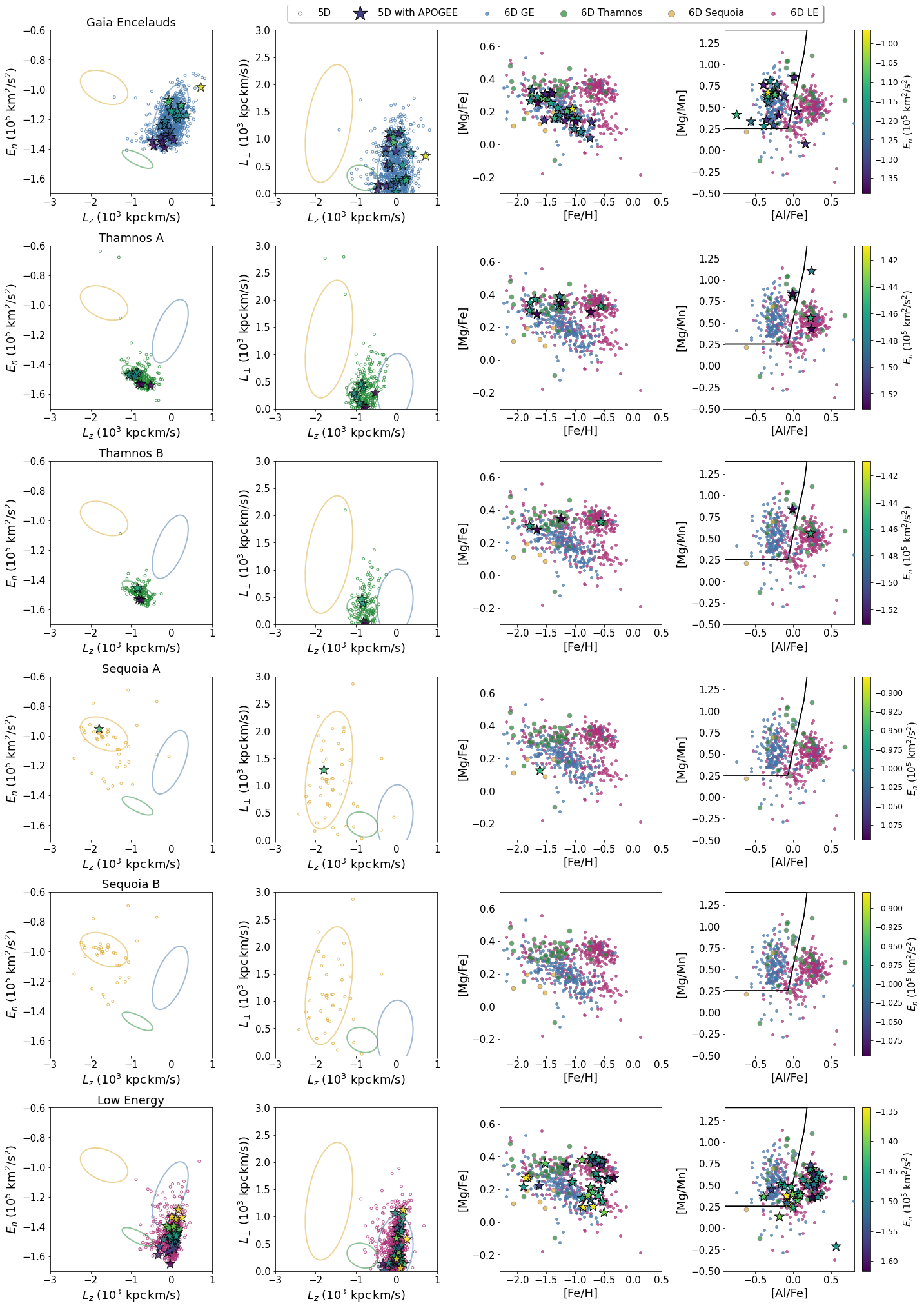}
\caption{ Left panels: Stars with radial velocities from the 5D samples (selected assuming $V_{\textrm{los}}$=0 km/s) and where they are distributed in IoM space using their RVs. We see that the 5D selected samples are more extended than the original definitions from \citet{Dodd2023} shown for GE, Thamnos, and Sequoia by the ellipsoids in the corresponding colours. Typically selecting stars to higher L$_\perp$. Stars with a star marker are highlighted since they have APOGEE abundances, and they are shown in the last two panels compared to the true 6D definitions. We see that the 5D selected samples follow the same distributions as the parent 6D sample in these spaces.
}
\label{fig:5D_check_iom}
\end{figure*}

Our 5D selection method could be adding more contamination to our samples. We can use the 6D stars that end up in our 5D selected samples to check that we are not adding noise to these samples. In Sect. \ref{sec:5D_checks} we have shown that the 6D stars we are adding by our selection do not belong to other sub-structures but general halo stars. Here we show that these general halo stars are consistent with the 6D sample we are considering, in terms of their metallicity distribution. Fig.~\ref{fig:5D_check} shows the LAMOST DR7 LRS metallicity distributions of stars in our 5D samples compared to our 6D samples of the same sub-structure. 
Any changes in our MDFs are due to adding contaminating stars but we can see that our MDFs resemble the 6D MDFs, with some difference in the shape for Sequoia that is due to the low number of stars. Stars which have a LAMOST DR7 LRS metallicity or APOGEE DR17 abundances must also have a radial velocity from the corresponding survey so we can also check where these added stars are in IoM space. We show in Fig.~\ref{fig:5D_check_iom} the distribution, in IoM space, of all 6D stars that enter our 5D selections in the first two columns. Star markers correspond to those with APOGEE DR17 abundances coloured by their $E_n$ and these are shown in the last two columns compared to the true 6D samples in the background. This shows us that the added stars are slightly more extended in IoM space than our true 6D sub-structure but chemically they are following the same sequences as the 6D selections.

\FloatBarrier
\section{Parameters for CMD fitting routine}\label{sec:cmd_fit_parameters}
Table \ref{tab:bundle} can be used for reproducing the fitting bundle that has been used for selecting the region in which we perform the CMD fitting. 
Table \ref{tab:seeds} shows the seed points used for the CMD fitting. 
\begin{table}[h!]
\centering
\caption{Vertices of the bundle used for defining the region within which we performed the CMD fitting.}
\begin{tabular}{cc}
BP-RP Colour & M$_G$ \\
\hline
\hline
0.388 & $-1.615$\\
1.112&  $-3.326$\\
1.953& $-3.056$\\
1.948& $-0.633$\\
1.452& 1.458\\
1.209& 4.611\\
0.126& 4.579\\
$-0.419$& 1.284\\
$-0.419$& $-3.563$\\
$-0.424$& $-3.595$\\
0.388 &$-1.615$\\
\end{tabular}
\label{tab:bundle}
\end{table}

\begin{table}[h!]
\centering
\caption{Age (Z) seed points used for CMD fitting.}
\begin{tabular}{c|cc}
Age (Gyr) & Z & [Fe/H] (dex) \\
\hline
\hline
0.08 & 0.0001 & $-2.498$ \\
0.126 & 0.00013
& $-2.4 $\\
0.192 & 0.00015
& $-2.302$ \\
0.262 & 0.00020 
& $-2.204$ \\
0.334 & 0.00025 
& $-2.106$ \\
0.404 & 0.00031
& $-2.008$ \\
0.469 & 0.00039
& $-1.91 $\\
0.532 & 0.00048
& $-1.812$ \\
0.596 & 0.00061
& $-1.714$ \\
0.656 & 0.00076
& $-1.616$ \\
0.718 & 0.00095
& $-1.518$ \\
0.784 & 0.00119
& $-1.42$ \\
0.857 & 0.00149
& $-1.322$ \\
0.938 & 0.00187
& $-1.224$ \\
1.028 & 0.00234
& $-1.126$ \\
1.128 & 0.00293
& $-1.028$ \\
1.244 & 0.00366
&$-0.93 $\\
1.391 & 0.00457
& $-0.832$ \\
1.576 & 0.00571
& $-0.734 $\\
1.81 &  0.00712
& $-0.636 $\\
2.066 & 0.00887
& $-0.538$ \\
2.337 & 0.01104
& $-0.44 $\\
2.609 & 0.01372
& $-0.342 $\\
2.882 & 0.01702
&$ -0.244 $\\
3.156 & 0.02105
& $-0.146 $\\
3.427 & 0.02595
& $-0.048 $\\
3.695 & 0.03187
& 0.05 \\
3.978 & 0.039 & 0.148 \\
4.272 &   &   \\
4.581 &   &   \\
4.946 &   &   \\
5.389 &   &   \\
5.858 &   &   \\
6.351 &   &   \\
6.861 &   &   \\
7.372 &   &   \\
7.883 &   &   \\
8.393 &   &   \\
8.904 &   &   \\
9.415 &   &   \\
9.925 &   &   \\
10.436 &   &   \\
10.947 &   &   \\
11.457 &   &   \\
11.968 &   &   \\
12.479 &   &   \\
12.989 &   &   \\
13.5 &   &   \\
\end{tabular}

\label{tab:seeds}
\end{table}

\FloatBarrier
\clearpage
\section{Mocks with contamination}\label{appendix:mocks}
Here, we show the output age-metallicity distributions of our mock with added contamination from the low-energy (LE) sample described in Sect. \ref{sec:mocks_contamination}. 

\begin{figure}[h]
\centering

\includegraphics[width=0.5\textwidth]{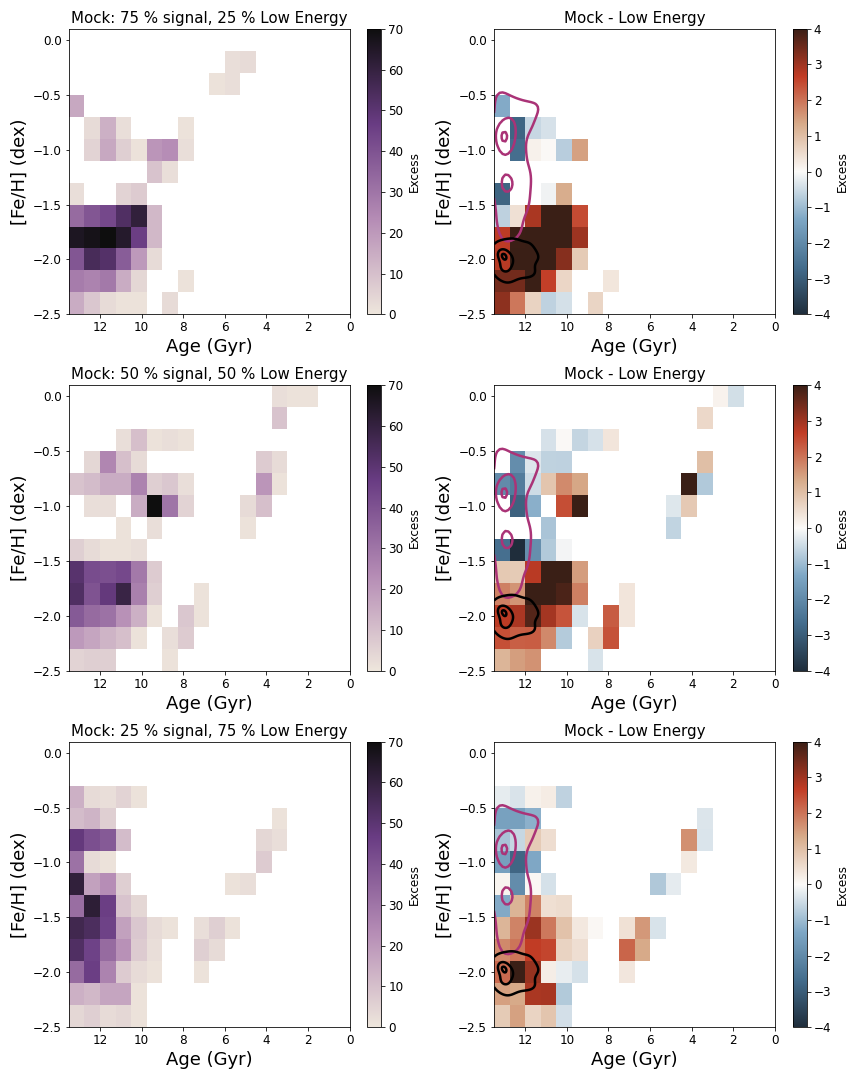}
\caption{Same as Fig.~\ref{toy_mocks:NS}, but for the case of the contamination from LE and the total number of stars being $N_T$. }
\label{toy_mocks:NT}
\end{figure}

\end{appendix}

\end{document}